 \documentclass[12pt, draftclsnofoot, journal, onecolumn]{IEEEtran}

\usepackage{cite}
\usepackage{multirow}		
\usepackage{subfigure}

 \usepackage[dvips]{graphicx}
 \graphicspath{{./}}
  \DeclareGraphicsExtensions{.eps, .jpg, .pdf, .png}
\usepackage{fancyhdr}

\ifCLASSINFOpdf
\else
\fi
\usepackage[cmex10]{amsmath}
\DeclareMathSizes{10}{10}{6.3}{4.5}
\usepackage{amssymb}

\usepackage{algorithm}              
\usepackage{algorithmic}   
\usepackage{enumitem}     
\usepackage{color}    

\makeatletter
\renewcommand{\maketag@@@}[1]{\hbox{\m@th\normalsize\normalfont#1}}%
\makeatother

\usepackage{amsfonts}
\begin{document}
%


\title{\huge Toward UL-DL Rate Balancing: Joint Resource Allocation and Hybrid-Mode Multiple Access for UAV-BS Assisted Communication Systems } 

%
%
%

\author{Haiyong Zeng,  Xu Zhu, \emph{Senior Member, IEEE}, Yufei Jiang, \emph{Member, IEEE},\\ Zhongxiang Wei, \emph{Member, IEEE}, and Sumei Sun, \emph{Fellow, IEEE}
 \thanks{
\IEEEpubidadjcol
Part of this work has been presented at IEEE Globecom 2020 \cite{Globecom2020}.

H.~Zeng, X.~Zhu and Y.~Jiang are with the School of Electronic and Information Engineering, Harbin Institute of Technology, Shenzhen, China. X.~Zhu is also with the School of Electrical Engineering and Electronics, University of Liverpool, UK.

Z.~Wei is with the College of Electronic and Information Engineering, Tongji University, Shanghai, China.

S.~Sun is with the Department of Communications and Networks, Institute for Infocomm Research, Agency for Science, Technology and Research, Singapore.
}
 }
\maketitle
\IEEEpeerreviewmaketitle


%
\begin{abstract}

In this paper, we investigate unmanned aerial vehicle (UAV) assisted communication systems that require quasi-balanced data rates in uplink (UL) and downlink (DL), as well as users' heterogeneous traffic. 
To the best of our knowledge, this is the first work to explicitly investigate joint UL-DL optimization for UAV assisted systems under heterogeneous requirements. A hybrid-mode multiple access (HMMA) scheme is proposed toward heterogeneous traffic, where non-orthogonal multiple access (NOMA) targets high average data rate, while orthogonal multiple access (OMA) aims to meet users' instantaneous rate demands by compensating for their rates.
HMMA enables a higher degree of freedom in multiple access and achieves a superior minimum average rate among users than the UAV assisted NOMA or OMA schemes.
Under HMMA, a joint UL-DL resource allocation algorithm is proposed with a closed-form optimal solution for UL/DL power allocation to achieve quasi-balanced average rates for UL and DL.
Furthermore, considering the error propagation in successive interference cancellation (SIC) of NOMA, an enhanced-HMMA scheme is proposed, which demonstrates high robustness against SIC error and a higher minimum average rate than the HMMA scheme.

\end{abstract}


\section{Introduction}

 Owing to high mobility, flexible deployment, low cost and high probability of providing line-of-sight (LoS) links, unmanned aerial vehicle (UAV) assisted communications is emerging as an effective complement to the fifth generation (5G) networks and has been widely deployed in various fast-growing applications, such as emergency communications, ubiquitous coverage, surveillance and monitoring \cite{Globecom2020, Al-Dhahir, Swindlehurst, Ng1}. As demonstrated in \cite{WuQQ, YCai, Ng2}, UAVs can be leveraged as aerial base stations (BSs) to significantly improve communication performance, \emph{e.g.}, providing higher data rate or expanding communication coverage, by establishing strong and flexible air-to-ground wireless communication connections with users \cite{YZeng, Zhao2,Schober}.

 The services provided by UAV-BS can be mainly classified into two types \cite{WuQQ3}: a) high rate-oriented services that require high average data rate, and b) instantaneous rate-sensitive services where a minimum instantaneous rate needs to be guaranteed at any time. 
In UAV assisted communications, it is challenging to satisfy the heterogeneous requirements of users at both uplink (UL) and downlink (DL). For example, in an open space stadium during a live event, the audiences may require video calls which are instantaneous rate-sensitive, while some other services, such as upload/download of captured image/video clip from the scene, are high rate-oriented. In addition, UL becomes increasingly important and even requires the same level of rate as DL in UAV communications \cite{Bergel1}\cite{Bergel2}, as exhibited in the live event example we have given. Most studies on conventional UAV assisted communications \cite{LBai,Senadhira,WuQQ2,SZhang} have focused on DL design only, leaving the achievable UL rate much lower than DL rate.
It is therefore timely and important to study on how to satisfy the heterogeneous traffic requirements and make a proper balance at both UL and DL in UAV assisted communications \cite{WuQQ3}\cite{WuQQ2}\cite{SZhang}.

\subsection{Related Work}
 A number of studies on UAV-BS assisted communications have been conducted \cite{WuQQ4,Lyu,WuQQ5,WuQQ6}. In \cite{WuQQ4}, UAV serves the ground users through time division multiple access (TDMA) scheme where at most one ground user can be served by UAV at each time slot. In \cite{Lyu}, the user scheduling and UAV trajectory design were jointly optimized to further enhance the minimum average rate among users. 
 In \cite{WuQQ5}, a UAV assisted orthogonal frequency division multiple access (OFDMA) system was investigated to meet the heterogeneous traffic demands of users, and the minimum  average rate among all users was maximized by joint bandwidth assignment, power allocation and UAV trajectory design. Nevertheless, the minimum average rate among users may decrease significantly as the number of users increases and it is difficult to guarantee all users' heterogeneous rate requirements, especially in some crowed hotspots.

In order to provide enhanced system spectrum efficiency (SE) performance and make full use of the underutilized wireless resources in orthogonality based approaches \cite{WuQQ4,Lyu,WuQQ5,WuQQ6}, power-domain non-orthogonal multiple access (NOMA) has been applied to UAV assisted systems, where a group of users share the same frequency band simultaneously with different power levels \cite{HaijunZhang,Liu,ZZhang,Sharma}. In \cite{Liu}, Liu \emph{et. al.} introduced the UAV assisted NOMA system and summarized their superiorities in terms of SE, multiple access, communication coverage, outage probability and energy efficiency (EE), over the UAV assisted orthogonal multiple access (OMA) methods.
 Although UAV assisted NOMA can provide a higher average rate, the instantaneous rate demands of users in each NOMA group may not be guaranteed due to the effects of 1) UAV movement; 2) the inter-user interference from other users within the same NOMA group; and 3) error propagation introduced by the imperfect successive interference cancellation (SIC) decoding process \cite{Sun,YZeng2,Zeng}. In this case, to guarantee the rate requirements of users, a hybrid OMA/NOMA terrestrial communication system was investigated in \cite{Marcano} and \cite{Song}, where dynamic mode switching between OMA and NOMA is utilized based on the instantaneous channel conditions. Nevertheless, the work in \cite{Marcano} and \cite{Song} for terrestrial communications cannot be applied to a UAV-BS assisted wireless system because of UAV's mobility and the high correlations between air-ground channels. Hence, to meet the heterogeneous traffic demands of ground users and consider fairness among users, it is significant to explore the UAV assisted hybrid non-orthogonal/orthogonal multiple access systems and its corresponding resource allocation.





In addition, as we have explained, UL plays an increasingly significant role in UAV assisted communications. It is therefore important to balance the UL/DL rates to meet the users' demands \cite{WuQQ3,Bergel1,Bergel2}. 
While most aforementioned studies regarding UAV assisted communications \cite{LBai,Senadhira,WuQQ2,SZhang} have focused on DL design only and lack of a consideration of joint DL and UL optimization. As a consequence, the achievable UL rate is generally much lower than DL rate, which makes it difficult to guarantee the increasing rate requirements of UL communication. 
In addition, in UL NOMA transmission, the power back-off strategy was adopted \cite{ZGDing} \cite{FFang}, where the users with weaker channel conditions were assigned less transmission power, as opposite to the power allocation principle of DL NOMA transmission\cite{Sun}. Hence, the work on DL NOMA or UL NOMA is not applicable to joint UL-DL optimization, and the joint UL-DL resource allocation for UAV assisted NOMA systems with UL and DL rate balancing remains an open challenge.

\subsection{Contributions}
Motivated by the above open issues, we propose a UAV assisted communication system with hybrid mode multiple access (HMMA) and joint UL-DL optimization, where both high rate-oriented and instantaneous rate-sensitive services are demanded simultaneously and the rates for UL and DL are required to be quasi-balanced. Our contributions are summarized as follows:

	1. To the best of our knowledge, this is the first explicit investigation of joint UL-DL optimization of bandwidth assignment, power allocation and trajectory design for the UAV assisted systems to achieve comparable UL-DL average rates and accommodate heterogeneous rate demands at UL and DL. Our work fills up the gap in current literature of UAV assisted communications \cite{LBai,Senadhira,WuQQ2,SZhang} which have focused on DL design only, leaving the UL rate much lower than DL rate and failing to guarantee the increasing rate requirements of UL communication.

	

	
2. We propose HMMA in which NOMA is used to achieve high average data rate, and OMA to meet users' instantaneous rate demands by compensating for the rates of users with poor channel conditions or strong inter-user interference. The proposed HMMA enables a higher degree of freedom in multiple access and allows dynamic resource allocation for each access mode. As a result, it is able to meet a wide range of heterogeneous traffic demands of users, and provide a superior minimum average rate performance and hence a better fairness among users, compared to the UAV assisted OMA \cite{WuQQ5} or NOMA \cite{Sun} approaches.

3. Making use of the reciprocity of the air-ground channel, we develop a joint UL-DL resource allocation (J-ULDL-RA) algorithm to conduct UL-DL bandwidth assignment, power allocation and UAV trajectory design in an alternative manner. In addition, by theoretically proving that the sum UL/DL transmission power is a convex function of UL/DL rates, the non-convex power allocation problem can be written as an equivalent convex form with respect to the rate of users, thus leading to the closed-form optimal power allocation solution. 

4. To mitigate the impact of the SIC error propagation in HMMA transmission, we propose an enhanced-HMMA (E-HMMA) scheme. The enhancement lies in that, in addition to the users suffering from poor channel conditions or strong inter-user interference, the users with severe error propagation caused by imperfect SIC are also taken into consideration for rate compensation by E-HMMA. The proposed E-HMMA scheme demonstrates higher robustness against the SIC error propagation and enables an enhanced minimum average rate than the HMMA scheme. Based on the proposed E-HMMA scheme, a joint UL-DL resource allocation algorithm with error propagation (J-ULDL-RAEP) is further developed for UAV assisted systems. Last but not least, it is also confirmed that both the proposed joint UL-DL optimization algorithms, namely J-ULDL-RA and J-ULDL-RAEP, can lead to fast convergence.

It is noteworthy that this paper is a substantial extension of our previous work in [1], as we propose a joint UL-DL resource allocation algorithm supported by a closed-form optimal solution to power allocation, whereas no power allocation solution was considered in [1]; also, a more generic system model with more than two users per NOMA group can be supported, as opposed to only two users per NOMA group in [1].

The rest of this paper is organized as follows. Section II introduces the UAV assisted HMMA system model and problem formulation. In Section III, the joint UL-DL resource allocation algorithm is proposed. The E-HMMA scheme is presented in Section VI. Section V presents the numerical results, followed by conclusion in Section IV.

	\begin{figure}[!t]
	\centering
	\includegraphics[width=5.2 in]{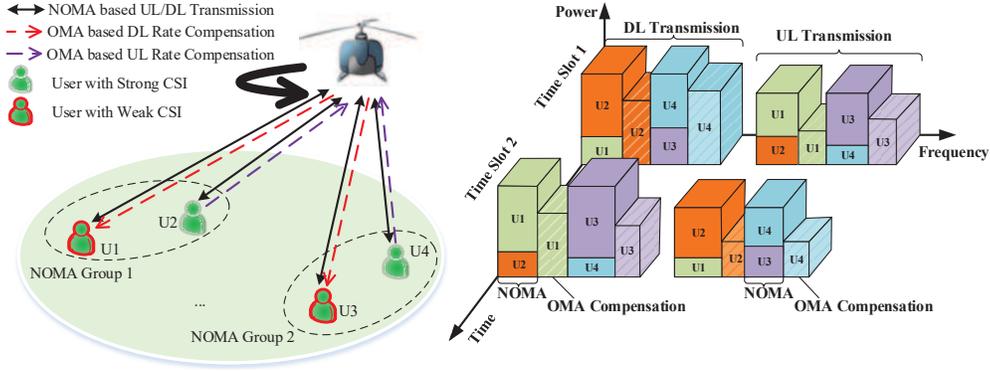}
	\caption{System model of UAV assisted HMMA systems, where both UL and DL communications are required to be guaranteed.}
\end{figure}

\section{ System Model}
 As illustrated in Fig. 1, we consider a UAV-BS assisted system with total bandwidth $B$ and $K$ ground users which can be divided into $M$ groups. 
 \textcolor{black}{The HMMA scheme is proposed to support the heterogeneous rate demands of users in both UL and DL, where NOMA is adopted by all users to communicate with UAV, and OMA is also utilized to help the users with low rate to meet their instantaneous rate demands.}
First, for each group, NOMA is utilized by the users within the group to communicate with the UAV-BS via distinct assigned UL or DL bandwidth. Note that although NOMA enables multi-user multiplexing and provides a high system rate \cite{HaijunZhang,Liu,ZZhang}, the instantaneous rate demands of some users in a NOMA group may not always be met due to UAV movement and the inter-user interference from other users within the same NOMA group, especially for the users with poor channel conditions (\emph{e.g.}, U1 and U3 in Fig. 1). To assure the instantaneous rate requirements of these users, the HMMA scheme provides, apart from NOMA transmission, a fraction of bandwidth to serve these users with OMA to boost up their rates. Denote the set of users in group $m$ $(m\in \{1,...,M\})$ as $\mathcal{K}_m$. 




Assume that the UAV flies above the ground users over a time duration of $T$, which can be discretized into $N$ equally spaced time slots. \textcolor{black}{Note that the value of $N$ for given $T$ should be carefully chosen so that the location of UAV within each time slot remains approximately unchanged as compared to the UAV-user link distance \cite{Zhao2}\cite{WuQQ5}.
Based on the discretization, the UAV's trajectory at time slot $n$ (${n\in \{ 1,2...,N\}}$) can be denoted as ${\bf{q}}[n]=[x[n],y[n]]$, where $[x[n],y[n]]$ are the UAV coordinates. By assuming that the UAV speed remains constant between any two time slots, a continuous UAV trajectory can be constructed by utilizing line-segments to connect the time slots in the discretized trajectory obtained by the proposed UAV trajectory design [20].}
Apparently, it needs to satisfy ${\rm{   }}||{\bf{q}}[n + 1] - {\bf{q}}[n]|| \le ({S_{{\rm{max}}}}T/N)^2$ for considering maximum speed ${S_{{\rm{max}}}}$ constraint, and guarantee ${\bf{q}}[N] = {\bf{q}}[1]$ for finishing a cyclic flight.
\textcolor{black}{In this paper, we assume that the UAV is deployed to serve the ground users in an open space such as outdoor sports events, where the channel state information (CSI) from the UAV to ground users is dominated by the LoS link and depends mainly on the distance between user and UAV. This is supported by the measurements in [33]-[35], where it is shown that at an altitude of more than 80 m, the probability of the presence of LoS from UAV to ground users in rural areas generally exceeds 95\% [35]. The same channel model has been utilized in  \cite{Zhao2}\cite{WuQQ4,Lyu,WuQQ5,WuQQ6}, \emph{etc.}} In addition, it is assumed that the UAV flies at a relatively high altitude $h$ and the Doppler effect of mobility has been perfectly compensated for \cite{Zhao2}\cite{WuQQ4}\cite{Lyu}. Hence, the channel power gain ${H_{k,n}}$ between user $k$ and UAV at time slot $n$ is ${H_{k,n}} = \frac{{{\gamma _0}}}{{{h^2} + ||{\bf{q}}[n] - {{\bf{w}}_k}|{|^2}}}$, with ${{\bf{w}}_k}$ and ${\gamma _0}$ respectively denoting the coordinate and channel power gain of user $k$ at the reference distance of ${d_0} = 1$ $\rm{m}$. 

In the following, the achievable rates of users at DL/UL transmission are respectively derived.




\subsection{Downlink HMMA Transmission}
Similar to \cite{HaijunZhang,Liu,ZZhang}, to make a balance between complexity and system performance, we assume the same time and frequency resource element can be shared by $L$ $(L\ge 2)$ users. For convenience, we assume the users are divided into $M$ groups with $L$ users in each group and keep the group fixed during the flight of UAV. 
For DL NOMA transmission, the UAV transmits the signals to $L$ users simultaneously over the same frequency resource \cite{Sun}, the received superposed signal ${D_{m,n,k}^\text{DL}}$ of user $k$ in NOMA group $m$ ($k \in \mathcal{K}_m$) at time slot $n$ can be expressed as
\begin{equation}
{D_{m,n,k}^\text{DL}}={\sqrt {{P_{m,n,k}^\text{NO,DL}}H_{k,n}}}{s_{m,n,k}^\text{DL}}+\sum\limits_{j \in {\mathcal{K}_m},j\ne k} {\sqrt {{P_{m,n,j}^\text{NO,DL}}H_{k,n}}} {s_{m,n,j}^\text{DL}}+z^\text{DL},
\end{equation}
where ${P_{m,n,k}^\text{NO,DL}}$ and ${s_{m,n,k}^\text{DL}}$ denote the transmission power and the signal of user $k$ at DL, respectively. $z^\text{DL}$ is the Gaussian noise with spectral power density $N_0$.


At the user end, SIC is adopted to decode the superposed signals. For DL NOMA transmission, the SIC decoding is generally in the increasing order of channel gains, \emph{i.e.}, the signal of the user with a weak channel gain (\emph{e.g.}, user $i$) is first decoded. Subsequently, the user with a higher channel gain (user $j$) removes the signal of user $i$ from the superposed signals and continues to decode its own signal \cite{HaijunZhang}\cite{Zeng}. Denote the bandwidth allocated to NOMA group $m$ for DL transmission as ${B_{m,n}^\text{{NO},DL}}$. 
As mentioned above, there are $L$ multiplexed users in each NOMA group, based on \cite{Sun}, to fairly compare the performance between NOMA and OMA under the same resources, a time slot in NOMA can be regarded as the combination of $L$ time slots in OMA. Hence, the rate of user $k$ for DL NOMA transmission at time slot $n$ can be obtained as
\begin{equation}
\begin{split}
R_{k,n}^{\text{NO,DL}} = L{B_{m,n}^{\text{NO,DL}}}{\log _2}\!\left(\!\!1 \!+\! \frac{{P_{m,n,k}^{\text{NO,DL}}H_{k,n}}}{\sum\limits_{j \in {\mathcal{K}^\text{DL}_{m,k}}} \!{{{P_{m,n,j}^\text{NO,DL}}H_{k,n}}\!+\! {N_0}{B_{m,n}^{\text{NO,DL}}}}}\!\!\right)\!,
\end{split}
\end{equation}
where ${\mathcal{K}^\text{DL}_{m,k}}$ denotes the set of users in NOMA group $m$ that have a higher channel gain than user $k$, \emph{i.e.}, $\mathcal{K}_{m,k}^\text{DL}=\left\{ {i}|{{H}_{{i},n}}\ge {{H}_{k,n}}\cap {i}\in \mathcal{K}_{m}^\text{DL} \right\}$. 




As mentioned above, though NOMA can provide a superior system performance via multi-user multiplexing, the instantaneous DL rate requirements of some users in NOMA group cannot always be guaranteed due to UAV movement and the inter-user interference introduced in NOMA transmission \cite{HaijunZhang,Liu,ZZhang}. To assure the DL instantaneous rate requirements of users and enhance user fairness, in our proposed HMMA, we allocate a fraction of bandwidth and transmission power apart from NOMA to serve these users with OMA to boost up their instantaneous rates. 
Denote the OMA bandwidth and transmission power for user $k$ in DL HMMA as ${B_{k,n}^\text{{O},DL}}$ and ${P_{k,n}^\text{{O},DL}}$, respectively. Specifically, ${B_{k,n}^\text{{O},DL}}=0$ indicates that user $k$ is allocated with no OMA bandwidth and requires no rate compensation in DL. The compensated rate $R_{k,n}^\text{{{O}},DL}$ of user $k$ can be obtained as 
\begin{equation}
R_{k,n}^{\text{{{O}},DL}} = {B_{k,n}^\text{{O},DL}}{\log _2}\left(1+\frac{{{P_{k,n}^\text{{{O}},DL}}{H_{k,n}}}}{{{N_{0}}{B_{k,n}^\text{{O},DL}}}}\right).
\end{equation}

The total DL rate of user $k$ at time slot $n$ is
\begin{equation}
R_{k,n}^\text{DL} =R_{k,n}^\text{{{NO}},DL} + R_{k,n}^\text{{{O}},DL}.
\end{equation}


\subsection{Uplink HMMA Transmission}
For UL transmission, the same user grouping is used as DL. \textcolor{black}{Note that due to the difference between UL and DL in NOMA, the HMMA scheme is operated differently in UL and DL. The differences between DL/UL NOMA lie in the following [23] [24] [28]. 1) Unlike the DL NOMA where UAV transmits the superposed signals to multiple users, in UL NOMA, the multiplexed users non-orthogonally transmit signals to the UAV over the same frequency resource element. 2) In DL NOMA, SIC is performed at the user side in the increasing order of users' channel gains [23] [28]. Whereas, in UL NOMA, SIC is conducted at the UAV side and the signal of a user with a higher channel gain is first decoded and removed from the superposed signals, so that the inter-user interference of weak users could be smaller. After that, the UAV continues to decode the signals of other users with weak channel gains \cite{ZZhang}.}

Based on the above, the received UL superposed signal $D_{m,n}^\text{UL}$ at UAV from the users in NOMA group $m$ can be expressed as 
\begin{equation}
{D_{m,n}^{\text{UL}}} = \sum\limits_{k \in {\mathcal{K}_m}} {\sqrt {{P_{m,n,k}^\text{NO,UL}}H_{k,n}}} {s_{m,n,k}^{\text{UL}}}  + z^\text{UL},
\end{equation}
where ${P_{m,n,k}^{\text{NO,UL}}}$ and ${s_{m,n,k}^{\text{UL}}}$ denote the UL transmission power and the signal of user $k$ at time slot $n$, respectively \cite{ZZhang}. 
Then the UAV decodes the superposed signals by the SIC approach. Denote the bandwidth allocated to group $m$ for UL NOMA transmission as ${B_{m,n}^{\text{NO},\text{UL}}}$, after SIC, the rate of user $k$ at UL NOMA transmission is given by
\begin{equation}
\label{RB6}
R_{k,n}^\text{NO,UL}=LB_{m,n}^\text{NO,UL}{{\log }_{2}}\left( 1+\frac{P_{m,n,k}^\text{NO,UL}H_{k,n}^{{}}}{\sum\limits_{j\in \mathcal{K}_{{m,k}}^\text{UL}}{P_{m,n,j}^\text{NO,UL}H_{j,n}^{{}}+{{N}_{0}}B_{m,n}^\text{NO,UL}}} \right),
\end{equation}
where ${\mathcal{K}^\text{UL}_{m,k}}$ denotes the set of users in ${\mathcal{K}^\text{UL}_{m}}$ that have lower channel gains than user $k$, \emph{i.e.}, $\mathcal{K}_{m,k}^\text{UL}=\left\{ {i}|{{H}_{{i},n}}\le {{H}_{k,n}}\cap {i}\in \mathcal{K}^\text{UL}_{m} \right\}$.

Based on (\ref{RB6}), the UL sum rate of users in NOMA group $m$ $R_{m,n}^\text{sum,UL}$ can be obtained as 
	\begin{equation}
	\begin{split}
	R_{m,n}^{\text{sum,UL}}=\sum\limits_{k\in {{\mathcal{K}}_{m}}}{R_{k,n}^\text{NO,UL}}=LB_{m,n}^\text{NO,UL}{{\log }_{2}}\left( 1+\frac{\sum\limits_{k\in {{\mathcal{K}}_{m}}}{P_{m,n,k}^\text{NO,UL}H_{k,n}^{{}}}}{{{N}_{0}}B_{m,n}^\text{NO,UL}} \right).
	\end{split}
	\end{equation}

It can be observed from (7) that the UL inter-user interference ${\sum\limits_{j\in \mathcal{K}_{{m,k}}^\text{UL}}{P_{m,n,j}^\text{NO,UL}H_{j,n}}}$ is naturally eliminated while deriving the UL sum rate of users, which can be utilized to help optimize UAV trajectory as detailed in Subsection III-B.



Similarly, to guarantee the instantaneous UL rate demands of users at each time slot, we allocate a fraction of bandwidth and transmission power to serve those users with poor channel conditions or severe inter-user interference using OMA. Then, the compensated rate of user $k$ for UL transmission at time slot $n$ can be expressed as
\begin{equation}
R_{k,n}^\text{{{O}},{UL}} ={B_{k,n}^\text{{O},{UL}}}{\log _2}\left(1 + \frac{{{P_{k,n}^\text{{{O}},{UL}}}{H_{k,n}}}}{{{N_{0}}{B_{k,n}^\text{{O},{UL}}}}}\right),
\end{equation}
where ${B_{k,n}^\text{{O},{UL}}}$ and ${P_{k,n}^\text{{O},{UL}}}$ denote the compensated UL bandwidth and transmission power of user $k$, respectively. Based on (6) and (8), the UL rate of user $k$ after rate compensation is
\begin{equation}
\label{HMMA9}
R_{k,n}^\text{UL}=R_{k,n}^\text{NO,UL}+R_{k,n}^\text{O,UL}.
\end{equation}



The total UL sum rate of users in NOMA group $m$ $R_{m,n}^\text{group,UL}$ can then be written as 
	\begin{equation}
\label{HMMA10}
	\begin{split}
	R_{m,n}^{\text{group,UL}} =R_{m,n}^{\text{sum,UL}}+\sum\limits_{k\in {{\mathcal{K}}_{m}}}{R_{k,n}^\text{O,UL}}.
	\end{split}
	\end{equation}

\subsection{ Problem Formulation}
We aim to maximize the minimum average rate among all users at both UL and DL transmissions by jointly optimizing UL-DL bandwidth assignment, power allocation and UAV trajectory design, while guaranteeing users' heterogeneous traffic demands. Specifically, we apply the concept of minimum-rate ratio (MRR) $\alpha$ for each user to indicate the ratio of the minimum data rate required (for instantaneous rate-sensitive traffic) over the total average rate (for both high rate-oriented and instantaneous rate-sensitive traffic) in $N$ time slots \cite{WuQQ5}, which specifies the percentage of the required instantaneous rate-sensitive traffic versus that of the high rate-oriented traffic, according to its applications, \emph{i.e.}, $\alpha$ fraction of its total average rate is instantaneous rate-sensitive and the remaining ($1-\alpha$) fraction is high average rate-oriented.
   
Denote the minimum average rate among users at UL and DL transmissions as $\eta \!=\!\mathop {\min }\limits_{i,k} \frac{1}{N}\!\!\sum\limits_{n = 1}^N \!\! {R_{k,n}^{i}}$, $k \in \left \{1,2,...,K\right\}$, with $i\in \{\text{UL,DL}\}$ denoting as the indicator for UL or DL. 
For bandwidth assignment and power allocation, let $\mathbf{B}^\text{DL}\!=\!\left\{[{B}_{m,n}^\text{NO,DL}]_{M\times N}, [{B}_{k,n}^\text{O,DL}]_{K\times N}\right\}$ and $\mathbf{B}^\text{UL}\!=\!\left\{[{B}_{m,n}^\text{NO,UL}]_{M\times N}, [{B}_{k,n}^\text{O,UL}]_{K\times N}\right\}$ denote the bandwidth assignment matrix, $\mathbf{P}^\text{DL}\!=\!\left\{[{P}_{m,n,l}^\text{NO,DL}]_{M\times N\times L},\right.$ $\left.[{P}_{k,n}^\text{O,DL}]_{K\times N}\right\}$, $\mathbf{P}^\text{UL}\!=\!\left\{[{P}_{m,n,l}^\text{NO,UL}]_{M\times N\times L}, [{P}_{k,n}^\text{O,UL}]_{K\times N}\right\}$ denote the power allocation matrix for DL and UL, respectively. Then, the overall bandwidth for DL and UL transmissions can be expressed as $B_n^\text{DL}=\sum\limits_{m = 1}^M {B_{m,n}^\text{NO,DL}}+\sum\limits_{k = 1}^K{{B_{k,n}^\text{O,DL}}}$ and $B_n^\text{UL}=\sum\limits_{m = 1}^M {B_{m,n}^\text{NO,UL}}+\sum\limits_{k = 1}^K{{B_{k,n}^\text{O,UL}}}$, respectively. Denote $\mathbf{Q}$ as the UAV trajectory matrix.
For UAV energy consumption, we consider two main components, namely, the communication related power ($P_t^\text{DL}$ and $P_t^\text{UL}$) and propulsion power $P(S_n)$. In particular, the propulsion power is consumed to keep UAV aloft and support its movement and can be approximated as a convex function with respect to UAV's flying speed $S_n$ \cite{YZeng}.


Then, with the consideration of heterogeneous rate requirements and UL-DL rate balancing, the optimization problem for the UAV assisted HMMA systems can be formulated as
 \begin{equation}
	\begin{split}
	\mathbf{P0:}\mathop {{\rm{max}}}\limits_{\mathbf{B}^\text{DL},\mathbf{B}^\text{UL},\mathbf{P}^\text{DL},\mathbf{P}^\text{UL},{\bf{Q}}} \eta 
	\end{split}
\end{equation}
 \[
 \begin{split}
	\text{s.t.}\quad&(11a): \frac{1}{N}\sum\limits_{n = 1}^N {R_{k,n}^{{{i}}} \ge } \eta^\text{}, {\rm{ }}\forall k,i,\notag\\
& (11b):{R_{k,n}^{{i}}} \ge \alpha_k \frac{1}{N}\sum\limits_{n = 1}^N {R_{k,n}^{{i}}}, \forall k,n,i,\notag\\
     &(11c):\sum\limits_{m = 1}^M\sum\limits_{l \in \mathcal{K}_m} {P_{m,n,l}^{\text{NO},i}}+\sum\limits_{k = 1}^K{P_{k,n}^{\text{O},i}}\le P_t^{i},\forall i,\notag\\ 
     &(11d):B_n^\text{DL}+B_n^\text{UL} \le B, \forall n,\notag\\
&(11e):P(S_n) \le P^\text{Pro}_{\max}, \forall n,\notag\\
     &(11f):||{\bf{q}}[n + 1] - {\bf{q}}[n]|| \le ({S_{{\rm{max}}}}T/N)^2, \notag\\
     &(11g):{\bf{q}}[N] = {\bf{q}}[1] ,\notag
\end{split}
\]
where (11a) guarantees the average rate of users at UL/DL transmission higher than a lower bound $\eta$, which is the objective to be optimized. (11b) denotes the UL/DL instantaneous rate requirements of users at each time slot. (11c) constrains the transmission power budget for DL and UL, respectively. (11d) indicates that the total utilizable bandwidth for UL and DL transmissions is limited to $B$. (11e) guarantees that the propulsion power of UAV should be not higher than the maximum propulsion power $P^\text{Pro}_{\max}$ in each time slot. Trajectory related constraints are confined by (11f) and (11g).

In its current form, problem $\mathbf{P0}$ is difficult to be solved directly. First, the rates ${R_{k,n}^{{\text{DL}}}}$ and ${R_{k,n}^{{\text{UL}}}}$ in constraints (11a)-(11b) are non-convex with respect to bandwidth assignment, power allocation and trajectory optimization variables ($\mathbf{B}^\text{DL}$, $\mathbf{B}^\text{UL}$, $\mathbf{P}^\text{DL}$, $\mathbf{P}^\text{UL}$ and ${\bf{Q}}$). Second, with a fixed UAV trajectory ${\mathbf{Q}}$, ${R_{k,n}^{{\text{DL}}}}$ and ${R_{k,n}^{{\text{UL}}}}$ are still non-convex due to the inter-user interference in the denominator of (3) and (6). Third, with a given bandwidth assignment and power allocation solution $\mathbf{B}^\text{DL}$, $\mathbf{B}^\text{UL}$, $\mathbf{P}^\text{DL}$, $\mathbf{P}^\text{UL}$, neither ${R_{k,n}^{{\text{DL}}}}$ nor ${R_{k,n}^{{\text{UL}}}}$ are concave or convex functions of ${\bf{Q}}$.
To address these challenges, we transform $\mathbf{P0}$ into the following
 \begin{equation}
	\mathbf{P1:}\mathop {{\rm{max}}}\limits_{\mathbf{B}^\text{DL},\mathbf{B}^\text{UL},\mathbf{P}^\text{DL},\mathbf{P}^\text{UL},{\bf{Q}}} \eta 
	 \end{equation}
	 \[
	 \begin{split}
	\text{s.t.}\quad&(11d)-(11g), \notag\\
&(12a):\frac{1}{N}\sum\limits_{n = 1}^N {R_{k,n}^{{\text{DL}}} \ge } \eta^\text{}, {\rm{ }}\forall k;\notag\\
&(12b):\frac{1}{N}\sum\limits_{n = 1}^N {R_{k,n}^{{\text{UL}}} \ge } \eta^\text{}, {\rm{ }}\forall k,\notag\\
     &(12c):{R_{k,n}^{\text{DL}}} \ge \alpha_k \eta, \forall k,n;\notag\\
&(12d):{R_{k,n}^{\text{UL}}} \ge \alpha_k \eta, \forall k,n,\notag\\
     &(12e):\sum\limits_{m = 1}^M \sum\limits_{l \in \mathcal{K}_m}{P_{m,n,l}^\text{NO,DL}}+\sum\limits_{k = 1}^K{P_{k,n}^\text{O,DL}} \le P_t^\text{DL},\notag\\
     &(12f):\sum\limits_{m = 1}^M\sum\limits_{l \in \mathcal{K}_m} {P_{m,n,l}^\text{NO,UL}}+\sum\limits_{k = 1}^K{P_{k,n}^\text{O,UL}}\le P_t^\text{UL}.\notag
	\end{split}
\]


Comparing (12c) and (12d) with (11b), it can be observed that the feasible solution of $\mathbf{P1}$ is generally a subset of that of $\mathbf{P0}$, and problems $\mathbf{P1}$ and $\mathbf{P0}$ are equivalent if all of the ground users can achieve equivalent average rate \cite{WuQQ5}. 
Now, problem $\mathbf{P1}$ is still non-convex. To this end, $\mathbf{P1}$ can be decomposed into several sub-problems, namely (1) bandwidth assignment and power allocation, and (2) UAV trajectory design, which can be alternatively solved by the proposed joint UL-DL resource allocation algorithm to obtain effective and low-complexity solutions, as detailed in Section III.







\section{HMMA and Joint UL-DL Resource Allocation}
In this section, an efficient J-ULDL-RA algorithm is proposed to jointly optimize the sub-problems in an alternative manner for UAV assisted HMMA systems.
The bandwidth assignment and power allocation are firstly optimized. Specifically, the bandwidth assignment is solved with given UAV trajectory and assuming the transmission power is equally distributed across the bandwidth. Afterwards, we convert the non-convex power allocation problem to a convex form with respect to users' rates by theoretically proving that the sum UL/DL transmission power is a convex function of UL/DL rates, hence obtaining the optimal power allocation. 
Finally, the successive convex optimization method is utilized to design the UAV trajectory.

\subsection{UL-DL Bandwidth Assignment and Power Allocation}
\subsubsection{UL-DL Bandwidth Assignment}
With a given UAV trajectory $\mathbf{Q}$, the bandwidth assignment and power allocation problem for UAV assisted HMMA systems can be formulated as
\[
\begin{array}{l}
\mathbf{P1.1:}{\rm{ }}\mathop {{\rm{max}}}\limits_{\mathbf{B}^\text{DL},\mathbf{B}^\text{UL}, \mathbf{P}^\text{DL},\mathbf{P}^\text{UL}} \;\;\;\;\;\eta \\
	\text{s.t.}\;\;\;(12a)-(12f) \;\text{and}\; (11d),
\end{array}
\]
 which is still non-convex with respect to ${\bf{B}^\text{DL},\bf{B}^\text{UL}}$ and ${\bf{P}^\text{DL},\bf{P}^\text{UL}}$. 
 In the following, \textbf{\emph{Proposition 1}} is presented to transform problem $\mathbf{P1.1}$ to a more feasible form.
 
\textbf{\emph{Proposition 1:}}
In UAV assisted HMMA systems, \textcolor{black}{given that the DL/UL transmission power is equally allocated over the bandwidth, \emph{i.e.},  $P_t^i/B_n^i=P_{m,n}^{\text{NO},i}/B_{m,n}^{\text{NO},i}=P_{m,n}^{\text{O},i}/B_{m,n}^{\text{O},i}$, $i\in \left \{\text{DL,UL}\right\}$,} problem $\mathbf{P1.1}$ can be transformed to a convex form as
\setcounter{equation}{12}
 \begin{equation}
	\mathbf{P1.2:}\mathop {{\rm{max}}}\limits_{\mathbf{B}^\text{DL},\mathbf{B}^\text{UL}} \eta  
 \end{equation}
	 \[
 \begin{split}
	    \text{s.t.}\quad	&(13a):\frac{1}{N}\sum\limits_{n = 1}^N {\tilde{R}_{k,n}^{{\text{DL}}} \ge } \eta^\text{}, {\rm{ }}\forall k;\;(13b):\frac{1}{N}\sum\limits_{n = 1}^N {\tilde{R}_{k,n}^{{\text{UL}}} \ge } \eta^\text{}, {\rm{ }}\forall k;\notag\\
	 &(13c):{\tilde{R}_{k,n}^{\text{DL}}} \ge \alpha_k \eta^{\text{}}, \forall k,n;\;(13d):{\tilde{R}_{k,n}^{\text{UL}}} \ge \alpha_k \eta^{\text{}}, \forall k,n;\;(13e):B_n^\text{DL}+B_n^\text{UL} \le B,  \forall n,\notag
	\end{split}
\]
where the rate related terms ${\tilde{R}_{k,n}^{{\text{DL}}}}$ and ${\tilde{R}_{k,n}^{\text{UL}}}$ are given in (36) and (37) and convex with respect to $\mathbf{B}^\text{DL}$ and $\mathbf{B}^\text{UL}$, respectively.

\emph{Proof:} Please see Appendix A. $\hfill\blacksquare$

According to \textbf{\emph{Proposition 1}}, the solution to the original bandwidth assignment problem $\mathbf{P1.1}$ can be obtained by solving the convex problem $\mathbf{P1.2}$ via standard convex optimization solvers.


Based on \textbf{\emph{Proposition1}}, a two-step UL-DL bandwidth assignment scheme is proposed. \textcolor{black}{Specifically, in Step 1, since the bandwidths of DL $B_{n}^{\text{DL}}$ and UL $B_{n}^{\text{UL}}$ are initially unknown, we assume the DL/UL transmission power is equally allocated over the whole bandwidth $B$ (\emph{i.e.}, $B_{n}^{\text{DL}}=B_{n}^{\text{UL}}=B$) and solve $\mathbf{P1}.\mathbf{2}$ to obtain an optimized bandwidth ${B^*}_{n}^{\text{DL}}$ and ${B^*}_{n}^{\text{UL}}$ at DL/UL transmission at each time slot. Afterwards, in Step 2, based on the optimized bandwidth (${B^*}_{n}^{\text{DL}}$, ${B^*}_{n}^{\text{UL}}$), we substitute $B_{n}^{\text{DL}}={B^*}_{n}^{\text{DL}}$ and $B_{n}^{\text{DL}}={B^*}_{n}^{\text{DL}}$ into (13a)-(13d) and solve $\mathbf{P1}.\mathbf{2}$ again to obtain the bandwidth assignment solution for NOMA and OMA transmission in both DL and UL links (\emph{i.e.}, $\mathbf{B}_{n}^{\text{DL}}$, $\mathbf{B}_{n}^{\text{UL}}$). Note that to assure the bandwidth assignment performance, the two-step scheme is performed at each iteration of the overall Algorithm.}

\subsubsection{UL-DL Power Allocation}
Now the bandwidth has been assigned, we then attempt to optimize the power allocation of users ${\mathbf{P}^\text{DL},\mathbf{P}^\text{UL}}$. For power allocation in UAV assisted HMMA systems, we need to optimize 1) the transmission power for NOMA and OMA transmission and 2) the power allocation among users in NOMA groups.
The power allocation problem can be expressed as
\[
	\begin{split}
	&\mathbf{P1.3:}\mathop {{\rm{max}}}\limits_{\mathbf{P}^\text{DL},\mathbf{P}^\text{UL}} \eta  \\
	    & \text{s.t.}\quad (12a)-(12f).
	\end{split}
\]

Since the DL/UL NOMA transmission is impaired by the inter-user interference, as described in (2) and (6), the UL/DL rates ${R_{k,n}^{{\text{UL}}}}$ and ${R_{k,n}^{{\text{DL}}}}$ in constraints (12a)-(12d) are non-convex functions of UL/DL transmission power, as a result, the power allocation problem $\mathbf{P1.3}$ is non-convex with respect the power allocation matrix ${\mathbf{P}^\text{DL},\mathbf{P}^\text{UL}}$ and difficult to solve in its current form.

\textcolor{black}{Note that though $\mathbf{P1}.\mathbf{3}$ is non-convex with respect to ${{\mathbf{P}}^{\text{DL}}}$ and ${{\mathbf{P}}^{\text{UL}}}$, the objective function and constraints (12a)-(12d) are obviously convex with respect to the UL/DL rates of users ${{\mathbf{r}}^{i}}=\left\{ {{[r_{k,n}^{\text{NO},i}]}_{K\times N}},{{[r_{k,n}^{\text{O},i}]}_{K\times N}} \right\}$, where $r_{k,n}^{\text{NO},i}=R_{k,n}^{\text{NO},i}/(LB_{m,n}^{\text{NO},i})$ and $r_{k,n}^{\text{O},i}=R_{k,n}^{\text{O},i}/B_{k,n}^{\text{NO},i}$. That is to say, if the sum UL/DL transmission power of users (\emph{i.e.}, constraints (12e) and (12f)) can be proved to be an equivalent convex function of ${{\mathbf{r}}^{i}}$, $\mathbf{P1.3}$ is convex with respect to the UL/DL rates of users ${{\mathbf{r}}^{i}}$, and thus the optimal power allocation solution can be obtained.} 

In the following, \textbf{\emph{Theorem 1}} is presented to prove that the sum UL/DL transmission power of users is convex with respect to ${{\mathbf{r}}^{i}}$, and the power allocation problem $\mathbf{P1}.\mathbf{3}$ can be written as an equivalent convex optimization problem of ${{\mathbf{r}}^{i}}$ \cite{dynamic}\cite{dynamic2}.



\textbf{\emph{Theorem 1:}}  
In UAV assisted HMMA systems, the optimal solution to power allocation can be obtained by solving the equivalent convex optimization problem with respect to $\mathbf{r}^{i}$ as
 \begin{equation}
\label{UAV19}
   \mathbf{P1.4:} \;\;\underset{{{\mathbf{r}}^\text{DL},{\mathbf{r}}^\text{UL}}}{{\max }}\, \eta
 \end{equation}
\[
\begin{split}
\text{s.t.}\quad& (12a)-(12d);\;\\
& (14a):\sum\limits_{m = 1}^M P_{m,n}^\text{DL}(\mathbf{r}^\text{DL})\le P_t^\text{DL};\;(14b):\sum\limits_{m = 1}^M P_{m,n}^\text{UL}(\mathbf{r}^\text{UL})\le P_t^\text{UL},\notag
	\end{split}
\]
where the sum transmission power of users at UL and DL $P_{m,n}^\text{DL}(\mathbf{r}^\text{DL})$ and $P_{m,n}^\text{UL}(\mathbf{r}^\text{UL})$ are respectively given in (\ref{HMMA17}) and (\ref{HMMA18}), with the underscript ${\psi }(k)$ denoting the permutation of channel gains, \emph{i.e.}, $H_{{{\psi }}(1),n}\ge ...\ge H_{{{\psi }}(k),n}\ge ...\ge H_{{{\psi }}(L),n}$.
	\begin{equation}
	\label{HMMA17}
		\begin{split}
   P_{m,n}^\text{DL}\left( {\mathbf{r}^\text{DL}} \right)=& \frac{{{N}_{0}}B_{m,n}^\text{NO,DL}}{H_{{{\psi }}(1),n}}{{2}^{\sum\nolimits_{k=1}^{L}{r_{{{\psi }}(k),n}^\text{NO,DL}}}}+\sum\limits_{l=2}^{L}{\left( \frac{{{N}_{0}}B_{m,n}^\text{NO,DL}}{H_{{{\psi }}(l),n}}-\frac{{{N}_{0}}B_{m,n}^\text{NO,DL}}{H_{{{\psi }}(l-1),n}} \right)}{{2}^{\sum\nolimits_{j=l}^{L}{r_{{{\psi }}(j),n}^\text{NO,DL}}}}\\
   &+\sum\limits_{k\in {{\mathcal{K}}_{m}}}{\frac{\left( {{2}^{r_{k,n}^\text{O,DL}}}-1 \right){{N}_{0}}B_{k,n}^\text{O,DL}}{H_{k,n}}}-\frac{{{N}_{0}}B_{m,n}^\text{NO,DL}}{H_{{{\psi }}(L),n}},
		\end{split}
	\end{equation}
	\begin{equation}
    \label{HMMA18}
	\begin{split}
   P_{m,n}^\text{UL}\left( {\mathbf{r}^\text{UL}} \right)=&\frac{{{N}_{0}}B_{m,n}^\text{NO,UL}}{H_{{{\psi }}(1),n}}{{2}^{\sum\nolimits_{k=1}^{L}{r_{{{\psi }}(k),n}^\text{NO,UL}}}}
  +\sum\limits_{l=2}^{L}{\left( \frac{{{N}_{0}}B_{m,n}^\text{NO,UL}}{H_{{{\psi }}(l),n}}-\frac{{{N}_{0}}B_{m,n}^\text{NO,UL}}{H_{{{\psi }}(l-1),n}} \right)}{{2}^{\sum\nolimits_{j=l}^{L}{r_{{{\psi }}(j),n}^\text{NO,UL}}}}\\
  &+ \sum\limits_{k\in {{\mathcal{K}}_{m}}}{\frac{\left( {{2}^{r_{k,n}^\text{O,UL}}}-1 \right){{N}_{0}}B_{k,n}^\text{O,UL}}{H_{k,n}}}-\frac{{{N}_{0}}B_{m,n}^\text{NO,UL}}{H_{{{\psi }}(L),n}}.
\end{split}
	\end{equation}
\emph{Proof:} Please see Appendix B. $\hfill\blacksquare$ 

\textcolor{black}{Based on \textbf{\emph{Theorem 1}}, the optimal data rates of users ${{\mathbf{r}}^{*}}^{\text{DL}},{{\mathbf{r}}^{*}}^{\text{UL}}$ can be obtained by solving $\mathbf{P1}.\mathbf{4}$, and then the optimal power allocation can be readily obtained by (17)-(19) without any                                                                    no approximation \cite{dynamic}\cite{dynamic2}.}
	\begin{equation}
	\label{UAVGBS37}
	\begin{split}
	{P^*}_{k,n}^{\text{O,}i}=\left( 2^{{r^*}_{k,n}^{\text{O},i}}-1 \right){N_0B_{m,n}^{\text{NO},i}}, i \in \{ \text{DL,UL} \}.
	\end{split}
	\end{equation}		
	\begin{equation}
	\label{UAVGBS39}
	\begin{split}
		{P^*}_{m,n,{{\psi }}(k)}^{\text{NO,UL}}\!=\!\frac{2^{\sum\limits_{i=k+1}^{L}\!{{r^*}_{m,n,{\psi }(i)}^\text{NO,UL}}} \! \left( 2^{{{r^*}_{m,n,{\psi }(k)}^\text{NO,UL}}}\!-\!1 \right)\!{N_0B_{m,n}^\text{NO,UL}}}{{H_{{{\psi }}(k),n}}}
	\end{split}
	\end{equation}	
	\begin{equation}
	\label{UAVGBS38}
	\begin{split}
	{P^*}_{m,n,\psi(k)}^\text{NO,DL}= \sum\limits_{i=1}^{k}{{P^*}_{m,n,{{\psi }}(i)}^{\text{NO,DL}}}-\sum\limits_{i=1}^{k-1}{{P^*}_{m,n,{{\psi }}(i)}^{\text{NO,DL}}},
	\end{split}
	\end{equation}		
where
 \[
 \begin{split}
 \sum\limits_{i=1}^{k}{{P^*}_{m,n,{{\psi }}(i)}^{\text{NO,UL}}}\!=\!\frac{{{N}_{0}}B_{m,n}^\text{NO,DL}}{H_{{{\psi }}(1),n}}{{2}^{\sum\nolimits_{i=1}^{k}{{r^*}_{{{\psi }}(i),n}^\text{NO,DL}}}}\!-\!\frac{{{N}_{0}}B_{m,n}^\text{NO,DL}}{H_{{{\psi }}(k),n}} \!+\! \sum\limits_{i=2}^{k}\!{\left( \frac{{{N}_{0}}B_{m,n}^\text{NO,DL}}{H_{{{\psi }}(i),n}^{{}}}\!-\!\frac{{{N}_{0}}B_{m,n}^\text{NO,DL}}{H_{{{\psi }}(i-1),n}} \right)}{{2}^{\sum\nolimits_{j=i}^{k}\!{{r^*}_{{{\psi }}(j),n}^\text{NO,DL}}}}.
 \end{split}
 \]

\subsection{ UAV Trajectory Design}
Now, bandwidth assignment and power allocation have been obtained in Subsection III-A, the UAV trajectory problem for the UAV assisted HMMA systems can be expressed as
\[
\begin{array}{l}
\mathbf{P1.5:}\mathop {{\rm{max}}}\limits_{\bf{Q}} \;\;\;\;\;\eta \\
	\text{s.t.}\;\;\;(11e)-(11g) \; \rm{and} \;(12a)-(12d),
\end{array}
\]
where successive convex optimization method is utilized to solve the non-convex problem $\mathbf{P1.5}$. 

 \textcolor{black}{Due to the difference between UL and DL HMMA transmission, the UL/DL inter-user interferences in (2) and (6) are different, which leads to different operations in UL/DL for the UAV trajectory design.} First, for DL transmission, though the NOMA/OMA rates of users ($R_{k,n}^{\text{NO,DL}}$, $R_{k,n}^{\text{O},\text{DL}}$ in (2) and (3)) are non-convex functions of $\mathbf{q}[n]$, they are evidently convex with respect to $||\mathbf{q}[n]-{{\mathbf{w}}_{k}}|{{|}^{2}}$. We can use the first-order Taylor's series expansion of $||\mathbf{q}[n]-{{\mathbf{w}}_{k}}|{{|}^{2}}$ as the lower bound of users' rates \cite{Zhao2}. The lower bound of the rate $R_{k,n}^{\text{DL,lb}}$ of user $k$ is given in (20), with $\phi _{k,n}^{r}=||{{\mathbf{q}}^{r}}[n]-{{\mathbf{w}}_{k}}|{{|}^{2}}$.


	\begin{equation}
	\label{HMMA22}
		\begin{split}
&   R_{{{k}},n}^{\text{DL,lb}}=G_{k,n}^{\text{DL}}+ \frac{-B_{k,n}^\text{O,DL}P_{k,n}^\text{O,DL}{{\gamma }_{0}}{{\log }_{2}}e \times \left( \Phi_{k,n}-\phi^r_{k,n} \right)}{\left( {{h}^{2}}+\phi^r_{k,n} \right)\left({{N}_{0}}B_{k,n}^\text{O,DL} \left( {{h}^{2}}+\phi^r_{k,n} \right)+P_{k,n}^\text{O,DL}{{\gamma }_{0}} \right)}+ \\ 
 & \frac{-LB_{m,n}^\text{NO,UL}P_{m,n,k}^\text{NO,DL}{{\gamma }_{0}}{{\log }_{2}}e \times \left( \Phi_{k,n}-\phi^r_{k,n} \right)}{\!\!\!\!\left(\!\!{{N}_{0}}B_{m,n}^\text{NO,DL}\!\left(\! {{h}^{2}}\!+\!\phi^r_{k,n} \!\right)\!+\!\sum\limits_{j \in \mathcal{K}_{m,k}^\text{DL}}\!\!\!\!{P_{m,n,j}^\text{NO,DL}} {{\gamma }_{0}}\! \!\right)\!\!\!\!\left(\! {{N}_{0}}B_{m,n}^\text{NO,DL}( {{h}^{2}}\!+\!\phi^r_{k,n} \!)\!+\!{\sum\limits_{j \in \mathcal{K}_{m,k}^\text{DL}}\!\!{P_{m,n,j}^\text{NO,DL}}{{\gamma }_{0}} \!+\!{P_{m,n,k}^\text{NO,DL}}}{{\gamma }_{0}}\! \right)}
		\end{split}
	\end{equation}
	\begin{equation}
	\begin{split}
   G_{k,n}^{\text{DL}}=&LB_{m,n}^\text{NO,DL}{{\log }_{2}}\!\!\left(\!\!1\!+\!\frac{P_{m,n,k}^\text{NO,DL}{{\gamma }_{0}}}{{{N}_{0}}B_{m,n}^\text{NO,DL}\!\left( \!{{h}^{2}}\!+\!\phi^r_{k,n} \right)\!+\!\sum\limits_{j \in \mathcal{K}_{m,k}^\text{DL}}\!\!{P_{m,n,j}^\text{NO,DL}}{{\gamma }_{0}} }\! \!\right)\!+\\
   &\!B_{k,n}^\text{O,DL}{{\log }_{2}}\!\left( \!1\!+\!\frac{P_{k,n}^\text{O,DL}{{\gamma }_{0}}/{{N}_{0}}/B_{m,n}^\text{O,DL}}{ {{h}^{2}}\!+\!\phi^r_{k,n}}\! \right).
\end{split}
	\end{equation}

 \textcolor{black}{On the other hand, for UL transmission, due to the inter-user interference of the user $k$ in the denominator of (6), the UL NOMA rate of user $k$ is non-convex with respect to the term $||\mathbf{q}[n]-{{\mathbf{w}}_{k}}|{{|}^{2}}$, which is different from the case at DL. As a result, the method utilized in DL NOMA is not applicable for UL NOMA. It is worth noting that the inter-user interference can be naturally eliminated when calculating $R_{m,n}^{\text{sum,UL}}$ in (7), the UL sum rate of users in each NOMA group, which makes $R_{m,n}^{\text{sum,UL}}$ convex with respect to $||\mathbf{q}[n]-{{\mathbf{w}}_{k}}|{{|}^{2}}$. As a result, we transform constraint (12b) to $({\text{24b}}):\frac{1}{NL}\sum\limits_{n=1}^{N}{(R_{m,n}^{\text{sum,UL}}+\sum\limits_{k\in {{\mathcal{K}}_{m}}}{R_{k,n}^{\text{O,UL}}})}\ge {{\eta }^{{}}},\forall k,m$, where both $R_{m,n}^{\text{sum,UL}}$ and $\sum\limits_{k\in {{\mathcal{K}}_{m}}}{R_{k,n}^{\text{O,UL}}}$ are convex with respect to ${{\Phi }_{k,n}}=||\mathbf{q}[n]-{{\mathbf{w}}_{k}}|{{|}^{2}}$.} In this way, the lower bound of the overall UL rate of users $R_{m,n}^{\text{group,UL,lb}}$ in NOMA group $m$ can be derived, which is given in (22).

	\begin{equation}
	\label{HMMA24}
		\begin{split}
   R_{m,n}^{\text{group,UL,lb}}&=G_{m,n}^{\text{group,UL}} -   LB_{m,n}^\text{NO,UL}\sum\limits_{k\in \mathcal{K}_m^\text{UL}}{\frac{{{\gamma }_{0}}{{\log }_{2}}e P_{m,n,k}^\text{NO,UL}/ \left( {h^2+\Phi_{k,n}} \right)^2}{{{N}_{0}}B_{m,n}^\text{NO,UL}+\sum\limits_{i\in \mathcal{K}_m^\text{UL}}{\frac{{{\gamma }_{0}}P_{m,n,i}^\text{NO,UL}}{{{h}^{2}}\!+\!\phi^r_{k,n}\! }}} \left(\Phi_{k,n}-\phi^r_{k,n}\right) }+  \\
 & \sum\limits_{k\in \mathcal{K}_m^\text{UL}}{\frac{-B_{k,n}^\text{O,UL}P_{k,n}^\text{O,UL}{{\gamma }_{0}}{{\log }_{2}}e \times \left( \Phi_{k,n}-\phi^r_{k,n} \right)}{\left( {{h}^{2}}+\phi^r_{k,n} \right)\left({{N}_{0}}B_{k,n}^\text{O,UL} \left( {{h}^{2}}+\phi^r_{k,n} \right)+P_{k,n}^\text{O,UL}{{\gamma }_{0}} \right)}}
		\end{split}
	\end{equation}	
	\begin{equation}
	\begin{split}
   G_{m,n}^{\text{group,UL}}\!=\!LB_{m,n}^\text{NO,UL}{{\log }_{2}}\!\left(\! 1\!+\!\sum\limits_{k\in \mathcal{K}_m^\text{UL}}\!\!
{\frac{{P_{m,n,k}^\text{NO,UL}}{{\gamma }_{0}}/{{N}_{0}}}{B_{m,n}^\text{NO,UL}\left( {{h}^{2}}+\phi^r_{k,n} \right)}} \!\right)\!+\!B_{k,n}^\text{O,UL}{{\log }_{2}}\!\left(\! 1\!+\!\frac{P_{k,n}^\text{O,UL}{{\gamma }_{0}}/{{N}_{0}}}{B_{k,n}^\text{O,UL}\left( {{h}^{2}}\!+\!\phi^r_{k,n} \right)} \!\right)\\
\end{split}
	\end{equation}



Based on the above transformations, the original trajectory problem can be approximated as

\begin{equation}
\mathbf{P1.6:}\;\;\mathop {{\rm{max}}}\limits_{\textbf{Q}} \;\;{\eta }
\end{equation}
\[
\begin{split}
\text{s.t.}\quad&(11e)-(11g);\;(24a):{\rm{   }}\frac{1}{N}\sum\limits_{n = 1}^N {R_{k,n}^{{\text{DL,lb}}} \ge } {\eta},\forall k,n;\;(24b):\frac{1}{LN}\sum\limits_{n = 1}^N {R_{m,n}^{\text{group,UL,lb}} \ge } {\eta},{\rm{ }}\forall m,n;\notag\\
     &(24c):{R_{k,n}^{\text{DL,lb}}} \ge \alpha_k \eta^\text{}, \forall k,n;\;(24d):\frac{1}{L}{R_{k,n}^{\text{group,UL,lb}}} \ge \alpha_k \eta^\text{}, \forall k,n,\notag
\end{split}
\]
which is now a convex quadratic problem and can be effectively solved by CVX. 

\begin{algorithm}[t] 
	\caption{ Joint UL-DL Resource Allocation Algorithm} 
	\begin{algorithmic}[1] 
		\REQUIRE  
		$K,H,S_{\max },\theta, B,N_0,{\gamma _0},T,N,\varepsilon,\eta_0$, $i_{\max}$;\\
		\ENSURE $ {\bf{B}^\text{DL}},{\bf{B}^\text{UL}},{\bf{P}^\text{DL}},{\bf{P}^\text{UL}},{\bf{Q}} $\\ 
		\STATE Initialize UAV trajectory ${{\mathbf{Q}}_0} = \{ {\mathbf{q}_0}[n],\forall n\} $, set $i=1$.
		\REPEAT
		\STATE Solve $\mathbf{P1.2}$ for given $\mathbf{Q}_{i-1}$, obtain ${\mathbf{B}_i^\text{DL}},\mathbf{B}^\text{UL}_i$. \textcolor{black}{After that, solve $\mathbf{P1.4}$ for given $\mathbf{Q}_{i-1}$ and ${\mathbf{B}_i^\text{DL}}, \mathbf{B}^\text{UL}_i$, obtain ${\mathbf{P}_i^\text{DL}}, \mathbf{P}^\text{UL}_i$ and $\eta_{i}^\text{}$.} 
		\STATE Solve $\mathbf{P1.6}$ for given ${\mathbf{B}_i^\text{DL}}$, $\mathbf{B}^\text{UL}_i$ and ${\mathbf{P}_i^\text{DL}}, \mathbf{P}^\text{UL}_i$, obtain $\mathbf{Q}_{i} $.
		\STATE $i=i+1$
		\UNTIL{$\eta_{i}^\text{}-\eta_{i-1}^\text{}<\varepsilon \; | | \; i>i_{\max}$}
		\RETURN ${\mathbf{B}_{i}^\text{DL}}, \mathbf{B}^\text{UL}_{i}, {\mathbf{P}_{i}^\text{DL}}, \mathbf{P}^\text{UL}_{i}, {\bf{Q}}_{i} $. 
	\end{algorithmic}
\end{algorithm}

\subsection{Overall Algorithm Description}
The J-DLUL-RA algorithm is given in Algorithm 1. \textcolor{black}{The bandwidth assignment and power allocation sub-problem in $\mathbf{P1.2}$ and $\mathbf{P1.4}$, and the trajectory design sub-problem in $\mathbf{P1.6}$ are alternately solved until convergence.}

In the following, we analyze the computational complexity of the proposed J-ULDL-RA algorithm. Since the bandwidth assignment problem $\mathbf{P1.2}$ is a linear program with respect to ${\mathbf{B}^\text{DL},\mathbf{B}^\text{UL}}$, the complexity of solving $\mathbf{P1.2}$ is in the order of $O(2K^3N^3)$. The complexity of obtaining the closed-form power allocation solution in $\mathbf{P1.4}$ is $O(KN)$. Note that in CVX, the interior-point method is generally employed to solve the optimization problems, the computational complexity of trajectory design in $\mathbf{P1.6}$ is on the basis of the complexity analysis of interior-point method to solve the corresponding convex conic problem \cite{WuQQ5}. Since $\mathbf{P1.6}$ mainly contains $K(1+L)/L$ second-order cone constraints with size $2N+1$, $K(1+L)/LN$ second-order cone constraints with size $3$,  $2N-1$ second-order cone constraints of size $4$, and one linear equality constraint with size $4$, referring to the analytical expression in \cite{WuQQ5}\cite{WeiTWC2021}\cite{Wang}, the total computational complexity of solving $\mathbf{P1.6}$ is $O(K^{\frac{3}{2}}N^{\frac{7}{2}})$.
Denote $i_{\max}$ as the number of iterations required in Algorithm 1 for convergence. \textcolor{black}{Since all problems $\mathbf{P1.2}$, $\mathbf{P1.4}$ and $\mathbf{P1.6}$ require to be solved in each iteration, whose complexities are $O(2K^3N^3)$, $O(KN)$ and $O(K^{\frac{3}{2}}N^{\frac{7}{2}})$, respectively. Given $i_{\max}$ to record the maximum allowable number of iterations, the whole computational complexity of Algorithm 1 can be obtained as $O(i_{\max}(2K^3N^3+KN+K^{\frac{3}{2}}N^{\frac{7}{2}}))$.}


\section{Enhanced-HMMA Scheme}
The HMMA scheme proposed in Section III can effectively improve the minimum average rate among users over the UAV assisted OMA/NOMA systems while guaranteeing heterogeneous traffic demands by compensating rates for the users suffering from weak channel conditions or severe inter-user interference. Nevertheless, in NOMA transmission, due to the limited computational capacity at users, there exists residual power in SIC process carried to the decoding of next level, which may cause error propagation \cite{ZZhang}\cite{Zeng}. In light of this, an E-HMMA scheme is proposed in this section to combat the effect of error propagation where the users with severe error propagation in SIC are also taken into account for rate compensation, so that the minimum average rate among users and the user fairness can be further improved over HMMA. 




\subsection{Description of the E-HMMA}
As mentioned in Section II, for DL NOMA SIC decoding, the superposed signals are generally decoded by users in the increasing order of channel gains \cite{HaijunZhang}\cite{Zeng}. Nevertheless, due to limited computational capacity, there exists common impairment among NOMA users in SIC process, which may cause error propagation to the next-level decoding \cite{ZZhang}\cite{Zeng}. 
Let $\omega$ denote the proportion of SIC residual power. Based on (2), the rate of user $k$ at DL NOMA transmission with imperfect SIC decoding can be obtained as
\begin{equation}
\label{HMMA28}
\begin{split}
&	\hat{R}_{k,n}^{\text{NO,DL}} = L{{B}_{m,n}^{\text{NO,DL}}}{\log _2}\!\!\left(\!\!1 \!+\! \frac{{{P}_{m,n,k}^{\text{NO,DL}}}}{\sum\limits_{i \in {\mathcal{K}^\text{DL}_{m,k}}} \!\!\!{{{{P}_{m,n,i}^\text{NO,DL}}}\!+\! \omega \!\! \sum\limits_{j \in \{\mathcal{K}^\text{DL}_m/{\mathcal{K}^\text{DL}_{m,k}}\}} \!\!{{{P}_{m,n,j}^\text{NO,DL}}}\!+\! {N_0}{{B}_{m,n}^{\text{NO,DL}}}/H_{k,n}}}\!\!\right)
\end{split}
\end{equation}
where $\omega \sum\limits_{j \in \{\mathcal{K}^\text{DL}_m/{\mathcal{K}^\text{DL}_{m,k}}\}} {{{P}_{m,n,j}^\text{NO,DL}}H_{k,n}}$ denotes the SIC residual power at user $k$. 




After that, E-HMMA scheme is conducted for rate compensation to achieve enhanced minimum average rate among users and a higher user fairness. E-HMMA differs from HMMA by also taking into account the error propagation effect for rate compensation in addition to the poor channel conditions or strong inter-user interference in HMMA. With the E-HMMA scheme, the compensated rate $\hat{R}_{k,n}^\text{{{O}},DL}$ of user $k$ can be obtained as
\begin{equation}
\label{HMMA29-1}
\begin{split}
\hat{R}_{k,n}^\text{O,DL} = \underbrace{{{B}_{k,n}^\text{{O},DL}}{\log _2}\left(1+\frac{{{{P}_{k,n}^\text{{{O}},DL}}{H_{k,n}}}}{{{N_{0}}{{B}_{k,n}^\text{{O},DL}}}}\right)}_{\text{For inter-user interference compensation}} + \underbrace{{{B}_{k,n}^\text{{OE},DL}}{\log _2}\left(1+\frac{{{{P}_{k,n}^\text{{{OE}},DL}}{H_{k,n}}}}{{{N_{0}}{{B}_{k,n}^\text{{OE},DL}}}}\right)}_{\text{For SIC error compensation}},
\end{split}
\end{equation}
where ${{B}_{k,n}^\text{{{O}},DL}}$, ${{P}_{k,n}^\text{{{O}},DL}}$ and ${{B}_{k,n}^\text{{{OE}},DL}}$, ${{P}_{k,n}^\text{{{OE}},DL}}$ denote the compensated bandwidth and transmission power of user $k$ in the E-HMMA scheme to combat the inter-user interference and SIC error propagation, respectively. For example, when $L=2$ with ${H_{1,n}}>{H_{2,n}}$, according to (\ref{HMMA28}), the strong user $1$ may suffer from severe SIC error propagation when decoding the signal of user $2$, which deteriorates the rate of user $1$. Therefore, the compensated bandwidths of user $1$ in the E-HMMA scheme are ${{B}_{1,n}^\text{{{OE}},DL}}\ge 0$ and ${{B}_{1,n}^\text{{{O}},DL}}=0$. On the other hand, since the weak user $2$ undergoes inter-user interference from user $1$, its compensated bandwidths can be ${{B}_{2,n}^\text{{{O}},DL}}\ge 0$ and ${{B}_{2,n}^\text{{{OE}},DL}}= 0$.

With the proposed E-HMMA scheme, the rate of user $k$ at time slot $n$ can be obtained as 
\begin{equation}
\label{HMMA29}
\begin{split}
&\hat{R}_{k,n}^\text{DL} =\hat{R}_{k,n}^\text{{{NO}},DL} + \hat{R}_{k,n}^\text{O,DL}.
\end{split}
\end{equation}


On the other hand, for UL NOMA transmission, since BSs can accommodate much higher computation and signal processing requirement, the effect of SIC residual power can be readily ignored and hence we assume a perfect SIC process at UAV-BS \cite{ZZhang}. As a result, the UL rate $\hat{R}_{k,n}^\text{UL}$ of user $k$ after rate compensation and the total sum rate of users $\hat{R}_{m,n}^\text{group,UL}$ in NOMA group $m$ can be respectively expressed as 
\begin{equation}
\label{HMMA30}
\begin{split}
& \hat{R}_{k,n}^\text{UL}\!=\!{{B}_{k,n}^\text{{O},{UL}}}{\log _2}\!\left(\!1 \!+\! \frac{{{{P}_{k,n}^\text{{{O}},{UL}}}{H_{k,n}}}}{{{N_{0}}{{B}_{k,n}^\text{{O},{UL}}}}}\!\right)\!+\!L{B}_{m,n}^\text{NO,UL}{{\log }_{2}}\!\left( \!1\!+\!\frac{{P}_{m,n,k}^\text{NO,UL}H_{k,n}^{{}}}{\sum\limits_{j\in \mathcal{K}_{{m,k}}^\text{UL}}{{P}_{m,n,j}^\text{NO,UL}H_{j,n}^{{}}+{{N}_{0}}{B}_{m,n}^\text{NO,UL}}} \!\right),
\end{split}
\end{equation}
	\begin{equation}
	\label{HMMA31}
	\begin{split}
	\hat{R}_{m,n}^{\text{group,UL}} =\sum\limits_{k\in {{\mathcal{K}}_{m}}}{{{B}_{k,n}^\text{{O},{UL}}}{\log _2}\left(1 + \frac{{{{P}_{k,n}^\text{{{O}},{UL}}}{H_{k,n}}}}{{{N_{0}}{{B}_{k,n}^\text{{O},{UL}}}}}\right)}+LB_{m,n}^\text{NO,UL}{{\log }_{2}}\left( 1+\frac{\sum\limits_{k\in {{\mathcal{K}}_{m}}}{P_{m,n,k}^\text{NO,UL}H_{k,n}^{{}}}}{{{N}_{0}}B_{m,n}^\text{NO,UL}} \right),
	\end{split}
	\end{equation}
where the inter-user interference is eliminated naturally while deriving $\hat{R}_{m,n}^\text{group,UL}$ in (\ref{HMMA31}).

Denote the minimum average rate among users for the UAV assisted E-HMMA systems as $\hat{\eta} {\rm{ = }}\mathop {\min }\limits_{i,k} \frac{1}{N}\sum\limits_{n = 1}^N  {\hat{R}_{k,n}^{i}}$, $k \in \left \{1,2,...,K\right\}, i\in \{\text{UL, DL}\}$. Base on the above analysis, the max-min problem with the consideration of NOMA error propagation can be formulated as
\begin{equation}
	\mathbf{P2:}\mathop {{\rm{max}}}\limits_{\mathbf{B}^\text{DL},\mathbf{B}^\text{UL},{\bf{Q}}} \hat{\eta}
	 \end{equation}
	 \[
\begin{split}
	\text{s.t.}\quad& (11c)-(11g); \;(30a):\frac{1}{N}\sum\limits_{n = 1}^N {\hat{R}_{k,n}^{{\text{DL}}} \ge } \hat{\eta}, {\rm{ }}\forall k;\;(30b):\frac{1}{N}\sum\limits_{n = 1}^N {\hat{R}_{k,n}^\text{UL} \ge }\hat{\eta} ,{\rm{ }}\forall k;\notag\\
	&(30c):{\hat{R}_{k,n}^{\text{DL}}} \ge \alpha_k \hat{\eta}, \forall k,n;\;(30d):{\hat{R}_{k,n}^{\text{UL}}} \ge \alpha_k \hat{\eta}, \forall k,n,\notag
	\end{split}
\]
where $\mathbf{B}^\text{DL}\!=\!\left\{[{B}_{m,n}^\text{NO,DL}]_{M\times N}, [{B}_{k,n}^\text{O,DL}]_{K\times N}, [{B}_{k,n}^\text{OE,DL}]_{K\times N}\right\}$ and $\mathbf{B}^\text{UL}\!=\!\left\{[{B}_{m,n}^\text{NO,UL}]_{M\times N}, [{B}_{k,n}^\text{O,UL}]_{K\times N}\right\}$ denote the bandwidth assignment matrix, and $\mathbf{Q}$ is the UAV trajectory matrix.
\textcolor{black}{As verified by the simulation results in Section V, power allocation makes a marginal contribution to the overall performance of HMMA, compared to bandwidth assignment and UAV trajectory design. Hence, in this section, regarding to the E-HMMA with SIC error propagation, we assume uniform power allocation across the whole bandwidth to achieve a sub-optimal performance, following \cite{ZZhang}, \cite{Sun} and \cite{WeiTWC2019}.}  


\subsection{Joint UL-DL Resource Allocation with Error Propagation }
Different from $\mathbf{P1}$ in Section II, $\mathbf{P2}$ takes into account error propagation effect due to imperfect SIC process in NOMA transmission, hence is more complicated and practical. In addition, the E-HMMA scheme is utilized in $\mathbf{P2}$ where apart from rate compensation for the users with poor channel condition or strong inter-user interference (\emph{i.e.}, the HMMA scheme), a number of bandwidth is assigned to combat the  SIC error propagation, which results in an enhanced minimum average rate among users and a higher user fairness.

Due to the inter-user interference and SIC residual power introduced by NOMA, $\mathbf{P2}$ is apparently non-convex with respect to $\mathbf{B}^\text{DL},\mathbf{B}^\text{UL}$ and ${\bf{Q}}$. Consequently, $\mathbf{P2}$ is decomposed into bandwidth assignment and trajectory design, and a J-ULDL-RAEP algorithm is developed to alternatively solve these sub-problems.



\subsubsection{UL-DL Bandwidth Assignment }
With a given UAV trajectory $\mathbf{Q}$, the bandwidth assignment can be formulated as
 \[
	\begin{split}
	&\mathbf{P2.1:}\mathop {{\rm{max}}}\limits_{\mathbf{B}^\text{DL},\mathbf{B}^\text{UL}} \hat{\eta} \notag \\
	    \text{s.t.}&\; (11d)\; \text{and}\; (30a)-(30d),\notag
	\end{split}
\]
which is non-convex due to the inter-user interference and SIC residual power in the denominators of ${\hat{R}_{k,n}^{\text{DL}}}$ and ${\hat{R}_{k,n}^{\text{UL}}}$ in (\ref{HMMA28}) and (\ref{HMMA30}), respectively. Note that the transmission power is equally allocated over the bandwidth, \emph{i.e.}, $P_t^i/B_n^i=P_{m,n}^{\text{NO},i}/B_{m,n}^{\text{NO},i}=P_{k,n}^{\text{O},i}/B_{k,n}^{\text{O},i}=P_{k,n}^{\text{OE,DL}}/B_{k,n}^{\text{OE,DL}},  i\in \left \{\text{DL,UL}\right\}$, substituting the above equations into the DL/UL rate of users in (\ref{HMMA29}) and (\ref{HMMA31}) yields
 \begin{equation}
 \begin{split}
	\hat{R}_{k,n}^{\text{DL}} =&\left({B_{k,n}^\text{{O},DL}+B_{k,n}^\text{{OE},DL}}\right){\log _2}\!\left(\!\!1\!+\!\frac{{P_t^\text{DL} {H_{k,n}}}}{{{N_{0}}B_n^\text{DL}}}\!\!\right)\!+\\
	&\!L{B_{m,n}^\text{{NO},DL}} {\log _2}\!\left(\!\!1 \!+\! \frac{\theta_k {P_t^{\text{DL}}}}{{\sum\limits_{i\in \mathcal{K}_{m,k}^\text{DL}}\!\!{\theta_i P_t^{\text{DL}}} \!+\! \omega \!\! \sum\limits_{j \in \{\mathcal{K}^\text{DL}_m/{\mathcal{K}^\text{DL}_{m,k}}\}} \!\!\!\!\theta_j{P_t^{\text{DL}}}\!+\! \frac{{N_0}B_n^\text{DL}}{H_{k,n}}}}\!\!\right),
\end{split}
\end{equation}
\begin{equation}
 \begin{split}
	\hat{R}_{k,n}^{\text{UL}} = &L{B_{m,n}^{\text{NO,UL}}}{\log _2}\left(1 + \frac{\theta_k {P_t^{\text{UL}}H_{k,n}}}{{\sum\limits_{i\in \mathcal{K}_{m,k}^\text{UL}}{\theta_i P_t^{\text{UL}}H_{i,n} } + {N_0}B_n^\text{UL}}}\right)+{B_{k,n}^\text{{O},UL}}{\log _2}\left(1+\frac{{P_t^\text{UL} {H_{k,n}}}}{{{N_{0}}B_n^\text{UL}}}\right),
\end{split}
\end{equation}
where the power allocation ratio $\theta_k$ of user $k$ is pre-set, and the rates of users under error propagation are convex with respect to $\mathbf{B}^\text{DL},\mathbf{B}^\text{UL}$, respectively. Similar to Section III-A, we first solve $\mathbf{P2.1}$ to obtain an optimized bandwidth at DL/UL transmission at each time slot (\emph{i.e.}, ${B^*}_{n}^{\text{DL}}$ and ${B^*}_{n}^{\text{UL}}$), by assuming the DL/UL transmission power is equally allocated over the whole bandwidth $B$. After that, we substitute ${B^*}_{n}^{\text{DL}}$ and ${B^*}_{n}^{\text{UL}}$ into (30a)-(30d) and solve $\mathbf{P2.1}$ again to further optimize the bandwidth assignment for DL/UL NOMA and OMA transmission $\mathbf{B}_{n}^{\text{DL}}$, $\mathbf{B}_{n}^{\text{UL}}$.



\subsubsection{ UAV Trajectory Design}
After the bandwidth assignment has been obtained, we aim to solve the UAV trajectory design problem. Compared with the previous trajectory design in Subsection III-B, the difference is that besides the inter-user interference, the effect of error propagation should also be considered while designing UAV trajectory. It can be learned from (\ref{HMMA28}) that though there exists inter-user interference and SIC residual power in the denominator of the DL NOMA rate of users, it is convex with respect to $||{\bf{q}}[n] - {{\bf{w}}_{k}}|{|^2}$, hence the lower bound of the rate $\hat{R}_{k,n}^{\text{DL,lb}}$ of user $k$ at DL can be approximated by its first-order Taylor's series expansion of (\ref{HMMA29}), which is given by
\begin{equation}
	\label{HMMA36}
		\begin{split}
   & \hat{R}_{{{k}},n}^{\text{DL,lb}}= \hat{G}_{k,n}^{\text{DL}}+ \frac{\left( {{\left\| {\mathbf{q}}[n]-{\mathbf{w}_{k,n}} \right\|}^{2}}-{{\left\| {\mathbf{q}^{r}}[n]-{\mathbf{w}_{k,n}} \right\|}^{2}} \right) \times \left(-B_{k,n}^\text{O,DL}P_{k,n}^\text{O,DL}{{\gamma }_{0}}{{\log }_{2}}e\right)}{\left( {{h}^{2}}+{{\left\| {\mathbf{q}^{r}}[n]-{\mathbf{w}_{k,n}} \right\|}^{2}} \right)\left({{N}_{0}}B_{k,n}^\text{O,DL} \left( {{h}^{2}}+{{\left\| {\mathbf{q}^{r}}[n]-{{w}_{k,n}} \right\|}^{2}} \right)+P_{k,n}^\text{O,DL}{{\gamma }_{0}} \right)}  \\ 
 & + \frac{\left( {{\left\| {\mathbf{q}}[n]-{\mathbf{w}_{k,n}} \right\|}^{2}}-{{\left\| {\mathbf{q}^{r}}[n]-{\mathbf{w}_{k,n}} \right\|}^{2}} \right)\times \left(-LB_{m,n}^\text{NO,UL}P_{m,n,k}^\text{NO,DL}{{\gamma }_{0}}{{\log }_{2}}e \right)}{\left( {{N}_{0}}B_{m,n}^\text{NO,DL}\left(\! {{h}^{2}}\!+\!{{\left\| {\mathbf{q}^{r}}[n]\!-\!{{w}_{k,n}} \right\|}^{2}}\! \right)\!+\!\sum\limits_{j \in \mathcal{K}_{m,k}^\text{DL}}{P_{m,n,j}^\text{NO,DL}} {{\gamma }_{0}}+ \omega \! \sum\limits_{i \in \{\mathcal{K}^\text{DL}_m/{\mathcal{K}^\text{DL}_{m,k}}\}} \!{{{P}_{m,n,i}^\text{NO,DL}}{{\gamma }_{0}}}\! \!\right)} \times\\ 
 &{\left(\!\! {{N}_{0}}B_{m,n}^\text{NO,DL}( {{h}^{2}}\!+\!{{\left\| {\mathbf{q}^{r}}[n]\!-\!{{w}_{k,n}} \right\|}^{2}})\!+\!\!{\sum\limits_{j \in \mathcal{K}_{m,k}^\text{DL}}\!\!{P_{m,n,j}^\text{NO,DL}}{{\gamma }_{0}} +{P_{m,n,k}^\text{NO,DL}}}{{\gamma }_{0}}+ \omega \!\!\! \sum\limits_{i \in \{\mathcal{K}^\text{DL}_m/{\mathcal{K}^\text{DL}_{m,k}}\}} \!\!\!{{{P}_{m,n,i}^\text{NO,DL}}{{\gamma }_{0}}}\!\! \right)}^{\!\!-1}
		\end{split}
	\end{equation}
	\begin{equation}
	\begin{split}
   & \hat{G}_{k,n}^{\text{DL}}\!=\!{{\log }_{2}}\!\!\left(\!\! 1\!+\!\frac{P_{m,n,k}^\text{NO,DL}{{\gamma }_{0}}}{{{N}_{0}}B_{m,n}^\text{NO,DL}\!( {{h}^{2}}\!+\!{{\left\| {\mathbf{q}^{r}}[n]\!-\!{{w}_{k,n}} \right\|}^{2}})\!+\!\!\sum\limits_{j \in \mathcal{K}_{m,k}^\text{DL}}\!\!\!{P_{m,n,j}^\text{NO,DL}}{{\gamma }_{0}}+ \omega \!\! \sum\limits_{i \in \{\mathcal{K}^\text{DL}_m/{\mathcal{K}^\text{DL}_{m,k}}\}} \!\!\!\!{{{P}_{m,n,i}^\text{NO,DL}}{{\gamma }_{0}}}\! } \right)\\
   &\;\;\;\;\;\;\;\;\;\;\times LB_{m,n}^\text{NO,DL}+B_{k,n}^\text{O,DL}{{\log }_{2}}\left( 1+\frac{P_{k,n}^\text{O,DL}{{\gamma }_{0}}/({{N}_{0}}B_{m,n}^\text{O,DL})}{ {{h}^{2}}+{{\left\| {\mathbf{q}^{r}}[n]-{{w}_{k,n}} \right\|}^{2}}} \right).
\end{split}
	\end{equation}

Due to the inter-user interference at UL NOMA transmission, the UL NOMA rate is not a convex function of $||{\bf{q}}[n] - {{\bf{w}}_{k}}|{|^2}$. Note that the inter-user interference can be eliminated while calculating the UL sum rate of users in each NOMA group in (\ref{HMMA31}), constraint (30b) can be transformed to $\widetilde{\text{(30b)}}: \frac{1}{LN}\sum\limits_{n = 1}^N {\hat{R}_{m,n}^{\text{group,UL}}} \ge \hat{\eta},\forall k,m$. In this way, the lower bound of sum rate of users in NOMA group $m$ $\hat{R}_{m,n}^{\text{group,UL,lb}}$ can also be obtained.



Based on the above, the UAV trajectory problem can be formulated as
\begin{equation}
\mathbf{P2.2:}\;\;\mathop {{\rm{max}}}\limits_{\textbf{Q}} \;\;\hat{\eta }
\end{equation}
\[
	\begin{split}
\text{s.t.}\quad&(11e)-(11g);\; \widetilde{\text{(30b)}};\;(35a):{\rm{   }}\frac{1}{N}\sum\limits_{n = 1}^N {\hat{R}_{k,n}^{{\text{DL,lb}}} \ge } \hat{\eta},\forall k,n;\notag\\
	&(35b):\frac{1}{L}{\hat{R}_{k,n}^{\text{group,UL,lb}}} \ge \alpha_k \hat{\eta}, \forall k,n;\;(35c):{\hat{R}_{k,n}^{\text{DL,lb}}} \ge \alpha_k \hat{\eta}, \forall k,n,\notag
\end{split}
\]
which is a convex quadratic problem now and can be effectively solved by CVX. 

By alternately solving bandwidth assignment in $\mathbf{P2.1}$ and trajectory design in $\mathbf{P2.2}$, the optimization problem for UAV assisted E-HMMA systems with error propagation can be solved, the procedure of the proposed J-ULDL-RAEP algorithm is similar to Algorithm 1. According to \cite{WuQQ5}\cite{WeiTWC2021}\cite{Wang}, the complexities of solving problems $\mathbf{P2.1}$ and $\mathbf{P2.2}$ are $O(2K^3N^3)$ and $O(K^{\frac{3}{2}}N^{\frac{7}{2}})$, respectively. As a result, the whole computational complexity is $O(i_{\max}(2K^3N^3+K^{\frac{3}{2}}N^{\frac{7}{2}}))$.

\section{Simulation Results}
In this section, numerical results are carried out to demonstrate the superiority of our proposed UAV assisted hybrid OMA/NOMA multiple access schemes. Users are distributed in an area of 1.5 km $\times$1.5 km. The altitude of UAV is set as $h=100$ m and the maximum speed of UAV is $50$ m/s \cite{Zhao2}. The UL/DL transmission power ${P}_{{t}}^{\text{UL}}$ and ${P}_{{t}}^{\text{DL}}$ are $30$ dBm and $23$ dBm, respectively. The signal-sided power spectral density of Gaussian noise is ${N_0}=-110$ dBm/Hz \cite{Zhao2}. Other parameters are set to as ${\gamma_0}=-50$ dB, $B=2$ MHz, $T=50$ s, $K=6$, $L=2$, $\varepsilon=10^{-3}$, $P^\text{pro}_{\max}=1$ kW and $\theta=0.2$ \cite{YZeng}\cite{Sun}. 
\textcolor{black}{In addition, due to the lack of related work on the UAV assisted OMA or NOMA communications with joint UL-DL resource optimization, we select the most closely related systems, \emph{i.e.}, the UAV assisted OMA only in \cite{WuQQ5} and NOMA only in \cite{Sun} with their dedicatedly designed joint UL-DL resource allocation algorithms as benchmarks.}

\begin{figure}[!h]
	\centering
	\includegraphics[width=3.5 in ]{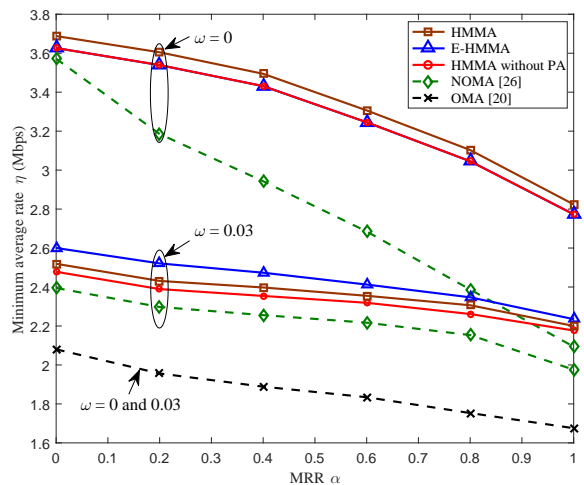}
	\caption{ \textcolor{black}{Minimum average rate $\eta$ for various UAV assisted systems with different values of MRR $\alpha$.}}
\end{figure}

 Fig. 2 illustrates the minimum average rate $\eta$ for various UAV assisted systems, with different values of MRR $\alpha$. 
As can be seen, $\eta$ decreases as $\alpha$ grows, and the proposed UAV assisted HMMA and E-HMMA schemes demonstrate a much higher minimum average rate than the UAV assisted NOMA or OMA approaches, especially when $\alpha$ reaches a relatively high level. For example, when $\alpha=0.8$ and $\omega=0$, the proposed HMMA and E-HMMA schemes achieve more than 27\% and 73\% higher minimum average rate over the UAV assisted NOMA and OMA schemes.
It is because that a dedicated fraction of bandwidth can be scheduled to compensate for the rates of NOMA users having poor performance, where the higher degree of freedom in dynamic resource allocation under each multiple access mode (OMA/NOMA) leads to a superior minimum average rate and higher user fairness. \textcolor{black}{In addition, the E-HMMA scheme outperforms HMMA when there exists error propagation in SIC decoding (\emph{e.g.}, $\omega=0.03$) since the effect of error propagation is captured in rate compensation by E-HMMA. With perfect SIC decoding (\emph{i.e.}, $\omega=0$), by jointly optimizing the power allocation for NOMA and OMA transmissions and the power allocation among users in each NOMA group, the HMMA scheme achieves a better minimum average rate than the HMMA without PA, which has the same performance of E-HMMA when $\omega=0$.}

\begin{figure}[!h]
	\centering
	\includegraphics[width=3.5 in]{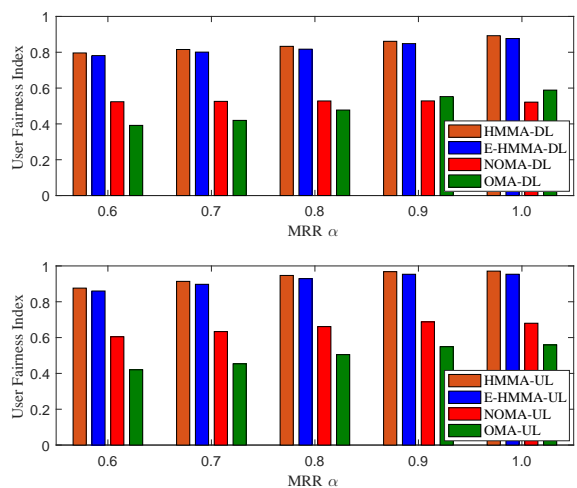}
	\caption{ \textcolor{black}{User fairness index for various UAV assisted communication systems, with perfect SIC decoding $\omega=0$. }}
\end{figure}

In Fig. 3, the user fairness index is presented to indicate the fairness among users under different values of MRR $\alpha$. For fair comparisons, the user fairness index considers both the minimum average rate $\eta$ and the instantaneous rates of users in different systems, which can be expressed as ${{J}^{j}}=\frac{{{\eta }^{j}}}{\max({{{\eta }^{j}}})}{\sum\limits_{n=1}^{N}{{{\left( \sum\limits_{k=1}^{K}{R_{k,n}^{{}}} \right)}^{2}}}}/\left({N\sum\limits_{n=1}^{N}{\sum\limits_{k=1}^{K}{R_{k,n}^{2}}}}\right)$, with $\frac{{{\eta }^{j}}}{\max({{{\eta }^{j}}})}$ denoting the normalized minimum average rate of mode $j$, $j\in \left\{ \text{HMMA,NOMA,OMA} \right\}$. As can be seen, the proposed HMMA and E-HMMA schemes demonstrate a higher user fairness over the UAV assisted NOMA or OMA methods, especially with stringent instantaneous rate demands. \textcolor{black}{Also, with perfect SIC decoding, the HMMA has a slightly higher user fairness than the E-HMMA scheme due to the utilization of PA in HMMA.} 



 \begin{figure}[!h]
	\centering
	\includegraphics[width=3.5 in ]{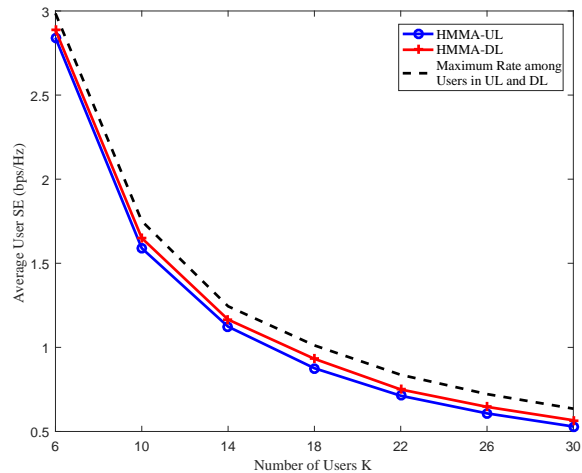}
	\caption{Impact of numbers of users $K$ on the average rate of users at UL and DL, with MRR $\alpha$ = 1, flight cycle $T$ = 150 s and SIC residual power proportion $\omega=0$.}
\end{figure}

The impact of different numbers of users $K$ on the UL/DL average rate $(\bar{R}^i =\frac{1}{KN}\sum\limits_{k= 1}^K{\sum\limits_{n = 1}^N  {R_{k,n}^{i}}}$, $i \in \left \{\text{UL,DL}\right\})$ of users is shown in Fig. 4. As can be seen, the average rate of users decreases with the number of users due to limited spectrum resources, and HMMA can achieve comparable average rates at UL and DL to meet the rate balancing requirement, which is more applicable to some crowded hotspots that require high rate at both links. In addition, both the curves of average rates in UL and DL are close to that of the maximum rate among users, which indicates the high user fairness of our proposed HMMA algorithm.

\begin{figure}[!h]
	\centering
	\includegraphics[width=3.5 in ]{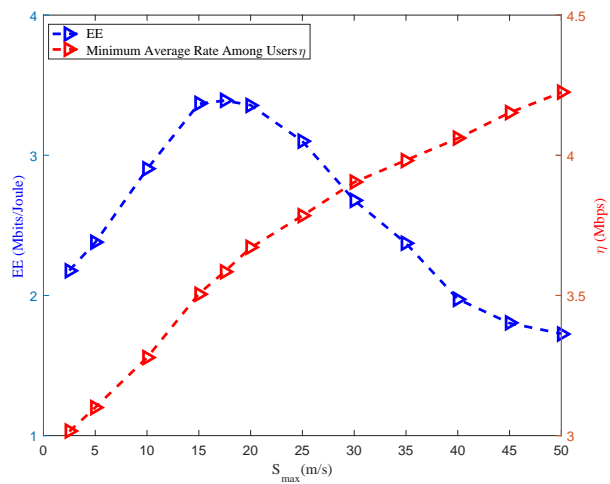}
	\caption{ Impact of maximum speed of UAV $S_{\max}$ on system EE performance and minimum average rate among users $\eta$, with MRR $\alpha=0$, flight cycle $T=125$ s and SIC residual power proportion $\omega=0$.}
\end{figure}


In Fig. 5, we evaluate the impact of maximum speed of UAV $S_{\max}$ on system EE and minimum average rate among users $\eta$, with $\alpha=0$, $T=125$ s and $\omega=0$. It can be learned from Fig. 5 that $\eta$ continues to increase as $S_{\max}$ becomes larger, whereas EE first increases as $S_{\max}$ grows and then decreases when it reaches a green point (\emph{i.e.}, $S_{\max}\approx 17.5$ m/s), which indicates that there is a tradeoff between $\eta$ and EE performance in UAV assisted HMMA systems.

\begin{figure}[!h]
	\centering
	\includegraphics[width=3.5 in ]{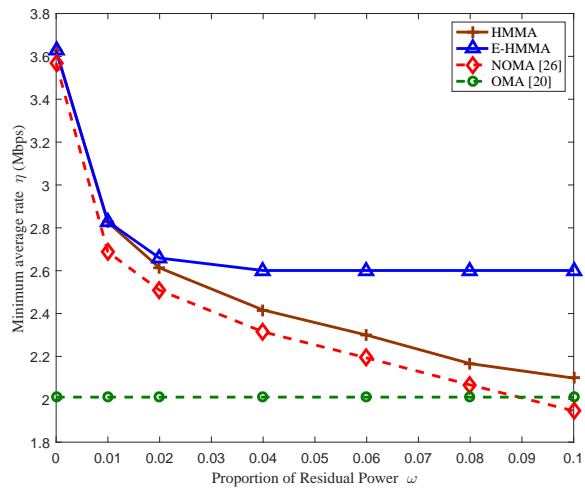}
	\caption{ Impact of NOMA error propagation on the minimum average rate among users $\eta$, with MRR $\alpha=0$.}
\end{figure}
\begin{figure}[!h]
	\centering
	\includegraphics[width=3.5 in ]{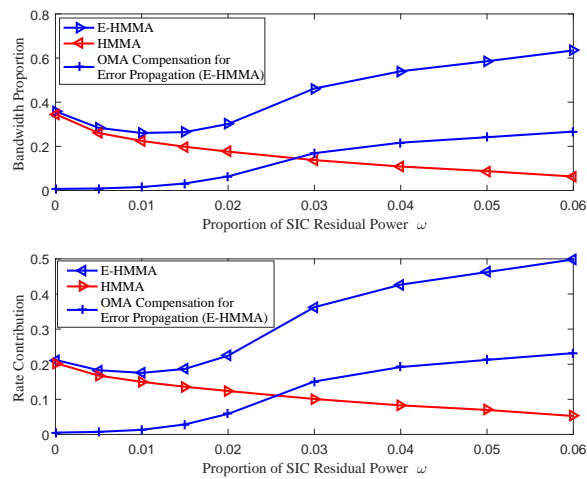}
	\caption{Bandwidth proportion and rate contribution for OMA compensation versus different values of SIC residual power proportion $\omega$, with MRR $\alpha=1$.}
\end{figure}

Fig. 6 shows the impact of SIC residual power $\omega$ on the minimum average rate of users, with $\alpha=0$. As can be seen, the proposed E-HMMA scheme is robust against NOMA error propagation, while the minimum average rates of UAV assisted HMMA and NOMA schemes decrease quickly as $\omega$ grows (\emph{i.e.}, the error propagation becomes severer). 
The reason is that, considering the effect of error propagation by E-HMMA, the rates of users with severe error propagation are also taken into consideration for rate compensation. In addition, with a larger $\omega$, more bandwidth and transmission power will be assigned to compensate for the rates of users suffering from severe error propagation, which makes the minimum average rate remain at a high level and contributes to a higher user fairness. 
The phenomenon can also be explained in Fig. 7, where the bandwidth proportion and rate contribution for OMA compensation in the proposed HMMA and E-HMMA schemes are compared, versus different values of proportion of SIC residual power $\omega$. As can be seen, when the error propagation ($\omega$) is small, the MRR requirement is the principal factor to have an influence on $\eta$, and the bandwidth proportions for OMA compensation in E-HMMA and HMMA are almost the same. It is because only the users with poor channel conditions or severe inter-user interference should be compensated for. As the NOMA error propagation becomes severer ($\omega$ grows), the E-HMMA scheme can allocate more bandwidth than HMMA to serve the users suffering from error propagation for rate compensation, so that the minimal average rate $\eta$ remains at a high level.  



\begin{figure}[!h]
	\centering
	\includegraphics[width=3.45 in]{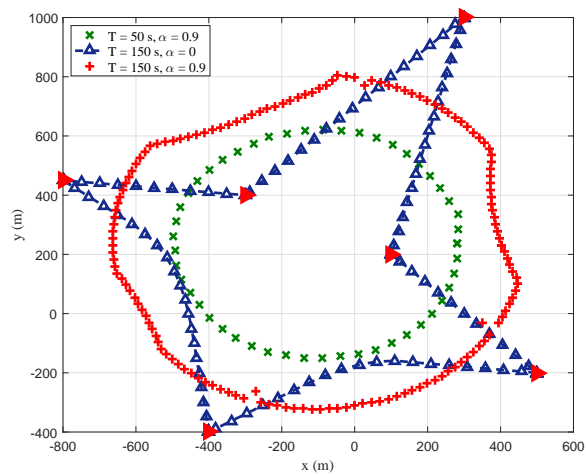}
	\caption{ UAV trajectories of the HMMA scheme with different MRR $\alpha$ and flight cycle $T$. }
\end{figure}

\begin{figure}[!h]
	\centering
	\includegraphics[width=3.5 in ]{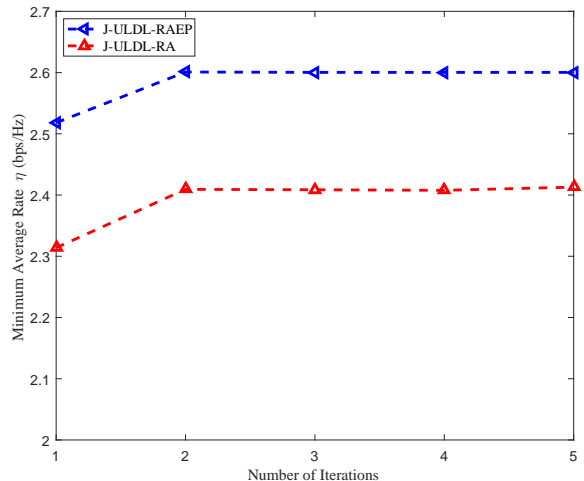}
	\caption{Convergence behavior of the proposed alternative algorithm, with accuracy factor $\varepsilon=10^{-3}$, MRR $\alpha=0$ and SIC residual power proportion $\omega=0.04$.}
\end{figure}

Fig. 8 illustrates the trajectories of the proposed UAV assisted HMMA systems, with different values of flight time $T$ and MRR $\alpha$. When the flight time is short, the UAV mainly flies in the central area due to maximum speed constraint. With a longer flight time, the UAV tends to fly closer to each user to enable a better LoS communication link. In addition, given a higher $\alpha$ (\emph{e.g.}, $T=150$ s and $\alpha=0.9$), the instantaneous rate requirements of users become stringent. As a result, the UAV adjusts its trajectory to fly around users so that the heterogeneous traffic demands of all users can be guaranteed.



Fig. 9 shows the convergence behavior of the proposed alternative algorithms. As can be seen, both the proposed joint ULDL resource allocation algorithms (J-ULDL-RA and J-ULDL-RAEP) require at most 3 iterations to convergence, confirming the low-complexity of the algorithms.


\section{Conclusion}
In this paper, we have proposed an HMMA scheme for UAV assisted systems where hybrid OMA/NOMA transmission is utilized to meet users' heterogeneous traffic demands as well as UL-DL rate balancing requirement. A J-ULDL-RA algorithm was developed for UL-DL bandwidth assignment, power allocation and UAV trajectory design. It has been demonstrated in simulation that, with a stringent instantaneous rate demand of users, the proposed HMMA achieves more than 27\% and 73\% higher minimum average rate than the UAV assisted NOMA \cite{Sun} and OMA \cite{WuQQ5} approaches, respectively. In addition, considering practical NOMA with imperfect SIC process, an E-HMMA scheme was further proposed, which is more robust against NOMA error propagation and demonstrates a higher minimum average rate and higher user fairness among users. Last but not least, both the proposed joint UL-DL resource allocation algorithms, namely J-ULDL-RA and J-ULDL-RAEP, can lead to fast convergence, proving their practicality in UAV assisted communications.


    \begin{appendices}

\section{Proof of Proposition 1}
Since the transmission power for both DL and UL is equally allocated over the bandwidth, we have $P_t^i/B_n^i=P_{m,n}^{\text{NO},i}/B_{m,n}^{\text{NO},i}=P_{m,n}^{\text{O},i}/B_{m,n}^{\text{O},i}$, $i\in \left \{\text{DL,UL}\right\}$, with $P_t^i$ denoting the transmission power for DL/UL transmission. 
Substituting the above equations into (4) and (9) yields
 \begin{equation}
 \begin{split}
	\tilde{R}_{k,n}^{\text{DL}} = &L{B_{k,n}^\text{{NO},DL}}{\log _2}\left(1 + \frac{\theta_k {P_t^{\text{DL}}H_{k,n}}}{{\sum\limits_{i\in \mathcal{K}_{m,k}^\text{DL}}{\theta_i P_t^{\text{DL}}H_{k,n} } + {N_0}B_n^\text{DL}}}\right)+{B_{k,n}^\text{{O},DL}}{\log _2}\left(1+\frac{{P_t^\text{DL} {H_{k,n}}}}{{{N_{0}}B_n^\text{DL}}}\right),
\end{split}
\end{equation}
\begin{equation}
 \begin{split}
	\tilde{R}_{k,n}^{\text{UL}} = &L{B_{k,n}^{\text{NO,UL}}}{\log _2}\left(1 + \frac{\theta_k {P_t^{\text{UL}}H_{k,n}}}{{\sum\limits_{i\in \mathcal{K}_{m,k}^\text{UL}}{\theta_i P_t^{\text{UL}}H_{i,n} } + {N_0}B_n^\text{UL}}}\right)+{B_{k,n}^\text{{O},UL}}{\log _2}\left(1+\frac{{P_t^\text{UL} {H_{k,n}}}}{{{N_{0}}B_n^\text{UL}}}\right),
\end{split}
\end{equation}
where $\theta_k$ denotes the power allocation ratio of user $k$ in NOMA group $m$, which is fixed and satisfies $\sum\limits_{k\in {{\mathcal{K}}_{m}}}\theta_k=1$. After such a transformation, the DL/UL rates of user $k$ in (36) and (37) are convex with respect to $\mathbf{B}^\text{DL}$ and $\mathbf{B}^\text{UL}$, respectively. As a result, we conclude that $\mathbf{P1.2}$ is convex. 

\section{Proof of Theorem 1}
First, for the downlink transmission, since there are $L$ users in each NOMA group $m$, the $L$ users are ranked based on the decreasing of channel gains,  $H_{{{\psi }}(1),n}\ge ...\ge H_{{{\psi }}(k),n}\ge H_{{{\psi }}(L),n}$.

According to (2), the transmission power to achieve $r_{m,n,{{\psi }}(k)}^\text{NO,DL}$ can be obtained as 
	\begin{equation}
	\label{UAVGBS18}
	\begin{split}
	P_{m,n,{{\psi }}(k)}^{\text{NO,DL}}=\left( 2^{r_{m,n,{{\psi }}(k)}^\text{NO,DL}}-1 \right)\left( {\sum\limits_{i=1}^{k-1}{P_{m,n,{{\psi }}(i)}^{\text{NO,DL}}}}+\frac{N_0B_{m,n}^\text{NO,DL}}{H_{{{\psi }}(k),n}}   \right).
	\end{split}
	\end{equation}
where ${\sum\limits_{i=1}^{k-1}{P_{m,n,{{\psi }}(i)}^{\text{NO,DL}}}}$ denotes the sum of transmission power for the first $(k-1)$ users. Note that the transmission power of user $k$ is determined by the first $k-1$ users, we define $S_k={\sum\limits_{i=1}^{k}{P_{m,n,{{\psi }}(i)}^{\text{NO,DL}}}}$, (\ref{UAVGBS18}) can be rewritten as
	\begin{equation}
	\label{UAVGBS19}
	\begin{split}
	S_k=\left( 2^{r_{m,n,{{\psi }}(k)}^\text{NO,DL}}-1 \right)\frac{N_0B_{m,n}^\text{NO,DL}}{H_{{{\psi }}(k),n}} + 2^{r_{m,n,{{\psi }}(k)}^\text{NO,DL}}S_{k-1}.
	\end{split}
	\end{equation}

Multiplying $D_k=2^{\sum\limits_{i=k+1}^{L}{r_{m,n,{{\psi }}(i)}^{\text{NO,DL}}}}$ at both sides of (\ref{UAVGBS19}) yields
	\begin{equation}
	\label{UAVGBS20}
	\begin{split}
	D_k S_k=D_{k-1} S_{k-1}+\frac{N_0B_{m,n}^\text{NO,DL}}{H_{{{\psi }}(k),n}}\left( D_{k-1}-D_{k} \right).
	\end{split}
	\end{equation}

Notice that $D_L=1$ and $S_0=0$, substituting into (\ref{UAVGBS20}), we have
	\begin{equation}
	\label{UAVGBS21}
	\begin{split}
	S_L=\sum\limits_{k=2}^{L}{\left( D_{k-1}-D_{k} \right)}\frac{N_0B_{m,n}^\text{NO,DL}}{H_{{{\psi }}(k),n}}.
	\end{split}
	\end{equation}
	
Consequently, the sum of DL transmission power in NOMA group $m$ can be expressed as
	\begin{equation}
	\label{UAVGBS22}
	\begin{split}
{P_{m,n}^{\text{NO,DL}}}\!=\!\frac{{{N}_{0}}B_{m,n}^\text{NO,DL}}{H_{{{\psi }}(1),n}}{{2}^{\sum\nolimits_{k=1}^{L}{r_{{{\psi }}(k),n}^\text{NO,DL}}}}\!-\!\frac{{{N}_{0}}B_{m,n}^\text{NO,DL}}{H_{{{\psi }}(L),n}} \!+\!\sum\limits_{k=2}^{L}{\left( \frac{{{N}_{0}}B_{m,n}^\text{NO,DL}}{H_{{{\psi }}(k),n}^{{}}}\!-\!\frac{{{N}_{0}}B_{m,n}^\text{NO,DL}}{H_{{{\psi }}(k-1),n}} \right)}{{2}^{\sum\nolimits_{j=k}^{L}{r_{{{\psi }}(j),n}^\text{NO,DL}}}}.
	\end{split}
	\end{equation}
 
Since $H_{{{\psi }}(k-1),n}\ge  H_{{{\psi }}(k),n}$, we have
	\begin{equation}
	\label{UAVGBS23}
	\begin{split}
 \frac{1}{H_{{{\psi }}(k-1),n}}\le  \frac{1}{H_{{{\psi }}(k),n}},
	\end{split}
	\end{equation}	
which implies that ${P_{m,n}^{\text{NO,DL}}}$ is convex with respect to $\mathbf{r}_n$ as it is a summation of non-negative convex functions.
The total DL NOMA transmission power can be obtained as
	\begin{equation}
	\label{UAVGBS23-1}
	\begin{split}
	P_{n}^{\text{NO,DL}}= \sum\limits_{m=1}^{M}{P_{m,n}^{\text{NO,DL}}}.
	\end{split}
	\end{equation}	
	
Then, for the downlink OMA transmission, based on (3), the transmission power to achieve $r_{k,n}^\text{O,DL}$ can be obtained as
	\begin{equation}
	\label{UAVGBS24}
	\begin{split}
	P_{k,n}^{\text{O,DL}}=\left( 2^{r_{k,n}^\text{O,DL}}-1 \right)\frac{N_0B_{k,n}^\text{O,DL}}{H_{k,n}}.
	\end{split}
	\end{equation}	


The DL OMA transmission power can be obtained as $P_{n}^{\text{O,DL}}= \sum\limits_{i=1}^{K}{P_{k,n}^{\text{O,DL}}}$. Hence, the total DL transmission power is
	\begin{equation}
	\label{UAVGBS26}
	\begin{split}
	P_{n}^\text{DL}(\mathbf{r}^\text{DL})=P_{n}^{\text{O,DL}}+P_{n}^{\text{NO,DL}}.
	\end{split}
	\end{equation}	

On the other hand, for UL transmission, we first obtain the UL OMA transmission power as
	\begin{equation}
	\label{UAVGBS27}
	\begin{split}
	P_{n}^{\text{O,UL}}= \sum\limits_{i=1}^{K}{\left( 2^{r_{k,n}^\text{O,UL}}-1 \right)\frac{N_0B_{k,n}^\text{O,UL}}{H_{k,n}}}.
	\end{split}
	\end{equation}	

Then we start to derive the sum power of UL NOMA transmission. According to (6), the UL transmission power to achieve $r_{m,n,{{\psi }}(k)}^\text{NO,UL}$ is given by
	\begin{equation}
	\label{UAVGBS28}
	\begin{split}
	P_{m,n,{{\psi }}(k)}^{\text{NO,UL}}=\left( 2^{r_{m,n,{{\psi }}(k)}^\text{NO,UL}}-1 \right)/{H_{{{\psi }}(k),n}} \times \left( {\sum\limits_{i=k+1}^{L}{P_{m,n,{{\psi }}(i)}^{\text{NO,UL}}{H_{{{\psi }}(i),n}}}}+{N_0B_{m,n}^\text{NO,UL}} \right).
	\end{split}
	\end{equation}
where ${\sum\limits_{i=k+1}^{L}{P_{m,n,{{\psi }}(i)}^{\text{NO,UL}}}}$ denotes the sum of transmission power for the first $(L-k-1)$ users whose signals can be firstly decoded in UL SIC process. Note that (\ref{UAVGBS28}) can be rewritten as 
	\begin{equation}
	\label{UAVGBS29}
	\begin{split}
	\sum\limits_{i=k}^{L}{H_{{{\psi }}(k),n}}P_{m,n,{{\psi }}(k)}^{\text{NO,UL}}=\left( 2^{r_{m,n,{{\psi }}(k)}^\text{NO,UL}}-1 \right){N_0B_{m,n}^\text{NO,UL}} + 2^{r_{m,n,{{\psi }}(k)}^\text{NO,UL}}\sum\limits_{i=k+1}^{L}{{H_{{{\psi }}(i),n}}P_{m,n,{{\psi }}(i)}^{\text{NO,UL}}}.
	\end{split}
	\end{equation}

Denote $F_k=\sum\limits_{i=k}^{L}{H_{{{\psi }}(i),n}}P_{m,n,{{\psi }}(i)}^{\text{NO,UL}}$ and $G_k=2^{\sum\limits_{i=1}^{k-1}{{r_{m,n,{{\psi }}(i),n}}}}$, multiplying $G_k$ at both sides of (\ref{UAVGBS29}) we have 
	\begin{equation}
	\label{UAVGBS30}
	\begin{split}
	F_k G_k=F_{k+1} G_{k+1}+{N_0B_{m,n}^\text{NO,UL}}\left( G_{k+1}-G_{k} \right).
	\end{split}
	\end{equation}

Note that $F_{(L+1)}=0$ and $G_1=1$, based on  (\ref{UAVGBS30}), we have
	\begin{equation}
	\label{UAVGBS31}
	\begin{split}
F_L G_L={\left( 2^{r_{m,n,{\psi }(L)}^\text{NO,UL}}-1 \right){N_0B_{m,n}^\text{NO,UL}}},
	\end{split}
	\end{equation}
	\begin{equation}
	\label{UAVGBS32}
	\begin{split}
		F_k=\sum\limits_{i=k}^{L}{H_{{{\psi }}(i),n}}P_{m,n,{{\psi }}(i)}^{\text{NO,UL}}=\frac{G_L-G_k}{G_k}N_0+F_L G_L=\left( 2^{\sum\limits_{i=k+1}^{L}{r_{m,n,{\psi }(i)}^\text{NO,UL}}}-1 \right){N_0B_{m,n}^\text{NO,UL}}.
	\end{split}
	\end{equation}

As a result, the transmission power of the $k$-th user $P_{m,n,{{\psi }}(k)}^{\text{NO,UL}}$ can be obtained as
	\begin{equation}
	\label{UAVGBS33}
	\begin{split}
		P_{m,n,{{\psi }}(k)}^{\text{NO,UL}}&=\frac{{F_{k}-F_{k+1}}}{{H_{{{\psi }}(k),n}}}=\frac{2^{\sum\limits_{i=k+1}^{L}{r_{m,n,{\psi }(i)}^\text{NO,UL}}} \left( 2^{{r_{m,n,{\psi }(k)}^\text{NO,UL}}}-1 \right){N_0B_{m,n}^\text{NO,UL}}}{{H_{{{\psi }}(k),n}}}.
	\end{split}
	\end{equation}

	The transmission power of users in NOMA group $m$ can be expressed as
	\begin{equation}
	\label{UAVGBS34}
	\begin{split}
		P_{m,n}^{\text{NO,UL}}\!=\!\frac{{{N}_{0}}B_{m,n}^\text{NO,UL}}{H_{{{\psi }}(1),n}}{{2}^{\sum\nolimits_{k=1}^{L}{r_{{{\psi }}(k),n}^\text{NO,UL}}}}\!-\!\frac{{{N}_{0}}B_{m,n}^\text{NO,UL}}{H_{{{\psi }}(L),n}}\!+\!\sum\limits_{k=2}^{L}\!\!{\left(\! \frac{{{N}_{0}}B_{m,n}^\text{NO,UL}}{H_{{{\psi }}(k),n}}\!-\!\frac{{{N}_{0}}B_{m,n}^\text{NO,UL}}{H_{{{\psi }}(k-1),n}} \!\right)}{{2}^{\sum\nolimits_{j=k}^{L}\!{r_{{{\psi }}(j),n}^\text{NO,UL}}}}.
	\end{split}
	\end{equation}
	
	According to (\ref{UAVGBS23}), $P_{m,n}^{\text{NO,UL}}$ is a summation of non-negative convex functions and is convex with respect to $\mathbf{r}_n$.
 The total DL NOMA transmission power is given by $P_{n}^{\text{NO,UL}}= \sum\limits_{m=1}^{M}{P_{m,n}^{\text{NO,UL}}}$.

As a result, the total UL transmission power is $P_{n}^\text{UL}(\mathbf{r}^\text{UL})=P_{n}^{\text{O,UL}}+P_{n}^{\text{NO,UL}}$, which is convex with respect to $\mathbf{r}_n$.

\end{appendices}

\end{document}